\documentclass{aa}  

\usepackage{graphicx}
\usepackage{rotating}
\usepackage{txfonts}
\usepackage[T1]{fontenc}
\usepackage{makecell}
\usepackage{multirow}
\usepackage{amssymb}
\usepackage{amsmath}
\usepackage{xcolor}
\usepackage{longtable}
\usepackage{pdflscape}

\usepackage[normalem]{ulem}

\newcommand{\hyperfootnote}[1][]{\def\ArgI\hyperfootnoteRelay}

\usepackage{hyperref}

\hypersetup{colorlinks=true, citecolor=blue, linkcolor=blue}

\begin{document} 

   \titlerunning{Spectroscopic AGN survey with NTT/SOFI for GRAVITY+ observations}
   \title{Spectroscopic AGN survey at $z$ $\sim$ 2 with NTT/SOFI for GRAVITY+ observations}
    \authorrunning{Santos, D. J. D.}
\author{D.~J.~D.~Santos\inst{1}
\and T.~Shimizu\inst{1}
\and R.~Davies\inst{1}
\and Y.~Cao\inst{1} 
\and J.~Dexter\inst{1,2}
\and P.~T.~de~Zeeuw\inst{3}
\and F.~Eisenhauer\inst{1,4}
\and N. M.~Förster Schreiber\inst{1}
\and H.~Feuchtgruber\inst{1}
\and R.~Genzel\inst{1,5} 
\and S.~Gillessen\inst{1}
\and L.~Kuhn\inst{1,6}
\and D.~Lutz\inst{1} 
\and T.~Ott\inst{1}
\and S.~Rabien\inst{1}
\and J.~Shangguan\inst{7,1}
\and E.~Sturm\inst{1} 
\and L. J.~Tacconi\inst{1}
}
\institute{Max Planck Institute for Extraterrestrial Physics (MPE), Giessenbachstr.1, 
85748 Garching, Germany
\and Department of Astrophysical \& Planetary Sciences, JILA, University of Colorado, Duane Physics Bldg., 2000 Colorado Ave, Boulder, CO 80309, US
\and Leiden University, 2311EZ Leiden, The Netherlands
\and Department of Physics, Technical University of Munich, 85748 Garching, Germany
\and Departments of Physics and Astronomy, Le Conte Hall, University of California,  Berkeley, CA 94720, USA
\and Department of Physics \& Astronomy, The University of British Columbia, Vancouver, Canada BC V6T 1Z1
\and Kavli Institute for Astronomy and Astrophysics, Peking University, Beijing 100871, People's Republic of China
}

   \date{Received xxxx xx, 2024; accepted xxxx xx, 2024}
 
\abstract{With the advent of GRAVITY+, the upgrade to the beam combiner GRAVITY at the Very Large Telescope Interferometer (VLTI), fainter and higher redshift active galactic nuclei (AGNs) are becoming observable, opening an unprecedented opportunity to further our understanding of the cosmic coevolution of supermassive black holes and their host galaxies. To identify an initial sample of high-redshift type~1 AGNs that can be observed with GRAVITY+, we have obtained spectroscopic data with NTT/SOFI of the most promising candidates. Our goal is to measure their broad line region (BLR) fluxes and assess their physical geometries by analysing the spectral profiles of their Balmer lines. We present 29 $z$ $\sim$ 2 targets with strong H$\alpha$ emission in the $K$-band. Their line profiles are strongly non-Gaussian, with a narrow core and broad wings. This can be explained as a combination of rotation and turbulence contributing to the total profile or two physically distinct inner and outer regions. We find small H$\alpha$ virial factors, which we attribute to the low full-width-half-maximum (FWHM)/$\sigma$ ratios of their non-Gaussian profiles, noting that this can lead to discrepancies in black hole masses derived from scaling relations. We also find two targets that show tentative evidence of BLRs dominated by radial motions. Lastly, we estimate the expected differential phase signals that will be seen with GRAVITY+, which will provide guidance for the observing strategy that will be adopted.}

   \maketitle
%
\section{Introduction}
\label{sec:intro}

Supermassive black holes (SMBHs) comprise the largest black holes in the Universe, with masses in the order of thousands to billions of solar masses. These black holes are believed to reside in the centre of galaxies and impact galaxy evolution through feedback mechanisms which regulate star formation and galaxy growth \citep{Silk2013, Terrazas2017, Harrison2017}. The growth of the host galaxy is also believed to affect the development of the central SMBH \citep{DiMatteo2005, Ferrarese2005, Peterson2008, Gultekin2009, Yesuf2020}. This almost symbiotic relationship between the central BH and its host galaxy is reflected in scaling relationships between the black hole mass and central stellar properties of the host galaxy \citep{Kormendy2013, Ferrarese2000, Marsden2020, MartinNavarro2020}. These findings invoke a coevolution scenario between the central BH and its host galaxy \citep{Heckman2014}.

Measuring SMBH masses is crucial to illuminating this scenario. In addition, measuring SMBH masses dynamically is essential to understand if and how the scaling relations typically used to infer those masses \citep{DallaBonta2020, Shen2012, Prieto2022} evolve with redshift.  
This is particularly important in early cosmic times when direct SMBH measurements are scarce.
Most SMBH mass measurements are done via spatially resolved stellar \citep{Saglia2016} or gas kinematics \citep{Boizelle2019, Osorno2023}, megamaser kinematics \citep{Greene2010, Kuo2020}, and reverberation mapping (RM) \citep{Peterson1993, Peterson2004, Li2023}.
However, these methods are predominantly used to probe SMBH masses of low redshift targets. Notably, performing RM at high redshifts requires long (multi-year) campaigns due to the cosmological time dilation and also because the quasars (QSOs) targeted tend to be luminous and so have large broad-line regions (BLRs) and, therefore, high SMBH masses. A recent summary of such efforts is given by \citet{Kaspi2021}.

Although several observational studies have already investigated the coevolution scenario at higher redshifts \citep[e.g.,][]{Lapi2014, Carraro2020}, these works use SMBH masses derived from scaling relations and so are based on the assumption that those relations, derived at low redshift, are also applicable at high redshift. As such, there is a need to measure SMBH masses at high redshift independently via direct dynamical methods.

With the unprecedented precision and resolution of GRAVITY \citep{GRAVITY2017}, the beam combiner at the Very Large Telescope Interferometer (VLTI), spatially resolving the broad line region (BLR) kinematics to measure SMBH masses has become possible in the local Universe \citep{GRAVITY2018, GRAVITY2020a, GRAVITY2021a, GRAVITY2024} with very long baseline interferometry \citep{Eisenhauser2023}. 
Extending this endeavour to higher redshifts was one of the imperatives for upgrading GRAVITY. 
The GRAVITY+ project aims to add wide-field off-axis fringe tracking (called GRAVITY-Wide) and new adaptive optics systems with laser guide stars (LGS) on all the unit telescopes (UTs). 
This will enable observations of both fainter and high-redshift quasars \citep{GRAVITY2022}. 
A key epoch to focus on is the ``cosmic noon'' in the redshift range 1 $<$ $z$ $<$ 3, corresponding to about 8-12 billion years ago when star formation and black hole growth both peaked \citep{Madau2014, Tacconi2020}. 
\cite{Abuter2024} recently performed the first dynamical mass measurement of a $z\sim$2 quasar with GRAVITY-Wide. With the other improvements from GRAVITY+, a vastly wider sky coverage will open up the possibility of selecting larger samples of AGNs and measuring their SMBH masses.
To prepare for this, we undertook a preparatory near-infrared (NIR) spectroscopic survey of promising AGN candidates for GRAVITY+.
This program aims to confirm suitable targets as quasars based on their $K$-band H$\alpha$ line profiles and fluxes. 
We identify the best targets for high-priority follow-up observations with GRAVITY+ by fitting their line profiles with a BLR model and estimating their expected interferometric signals.
This also yields information about their BLR geometries, giving us a first glimpse of what we can learn from their BLR structure.
We present 29 high-redshift targets ($z$ $\sim$ 2) observed in this initial survey. For our analyses, we adopt a Lambda Cold Dark Matter ($\Lambda$CDM) cosmology with $\Omega_m$ = 0.308, $\Omega_\Lambda$ = 0.692, and $H_0$ = 67.8 km s$^{-1}$ Mpc$^{-1}$ \citep{Planck2016}. 
We describe our sample selection and observations in Sect.~\ref{sec:targets_and_observations}.
We discuss our data reduction methods in Sect.~\ref{sec:data_reduction}. 
We summarise the emission line properties of our targets in Sect.~\ref{sec:emission_line_prop}. 
We discuss the BLR model used in this work, our fitting methodology and results in Sect.~\ref{sec:BLR_modelling}. 
We present the estimated differential phase signals of our targets based on our BLR model fitting in Sect.~\ref{sec:visphi_estimation}. 
Finally, we present our conclusions in Sect.~\ref{sec:conclusion}.
 
\section{Targets and observations}
\label{sec:targets_and_observations}

The targets were selected from the Million Quasar (Milliquas) Catalogue \citep{Flesch2021} version 7.5 (updated last 30th Apr. 2022) which provides a catalogue of type 1 QSOs and active galactic nuclei (AGNs) complete from literature, as well as from the Quaia spectroscopic catalogue \citep{StoreyFisher2024} that is based on Gaia candidates with unWISE infrared data.
Our selections of type~1 QSOs required 
(i) 2.1 $\lesssim z \lesssim 2.6$ so that H$\alpha$ is redshifted into the $K$-band, 
(ii) enabling GRAVITY+ observations either on-axis (i.e., $K < 13$) or off-axis (i.e., $K < 16$ as well as with a $K < 13$ star within 20$^{\prime\prime}$), and 
(iii) an initial prediction of the differential phase signal $>$ 0.3$^\circ$ to ensure that the integration times with GRAVITY+ would be reasonable. The differential phase is an interferometric observable that measures the centroid position as a function of wavelength (see Sect. \ref{sec:visphi_estimation} and \citealt{GRAVITY2020a} for more details). The expected differential phase signal was estimated by assuming all targets lie on the \citet{Bentz2013} R--L relation. The 5100\AA{} luminosity was estimated by scaling the mid-IR luminosity \citep{Krawczyk13} SED to the observed \textit{Gaia} $G_{\rm RP}$ magnitude. The phase signal was then calculated using $\Delta\phi = [f/(f + 1)]RB$ where $f$ is the typical line-to-continuum ratio of 3 for H$\alpha$, $R$ is the BLR radius (in radians), and $B$ is the projected baseline length of 100m for the VLTI divided by the wavelength. To make an initial assessment of the properties of the selected quasars, we also looked into the archival data, particularly within the Sloan Digital Sky Survey (SDSS) Data Release 16 Quasar Catalogue (DR16Q; \citealt{Lyke2020}) and the UV Bright Quasar Survey Catalogue (UVQS; \citealt{Monroe2016}) for their optical and ultraviolet (UV) spectra, respectively, to confirm AGN features such as the blue continuum and broad CIV, Mg{\small II}, or Ly$\alpha$ lines.

After the sample selection, 72 observable unique targets were selected. Observations were performed between April 2022 and February 2023 with the infrared spectrograph Son of ISAAC (SOFI) at the 3.6-m New Technology Telescope (NTT), which is operated by the European Southern Observatory (ESO). Due to weather conditions, we were able to observe 49 targets with NTT/SOFI. Since our goal is to pick objects that are suitable for GRAVITY+ observations, we only analysed those with sufficiently strong H$\alpha$ lines (i.e. integrated flux $\gtrsim$ 5 $\times$ 10$^{-16}$ erg s$^{-1}$ cm$^{-2}$). We then narrowed our sample to 29 targets. Some of the faint targets have signal-to-noise ratio (SNR) values between 4 and 10, while the brightest targets have SNR as high as 40-55. Table \ref{tab:targets} lists the 29 targets with good spectra, which are analysed in this work, together with a summary of their observations, which include the date of observation, exposure time, airmass, seeing, and SNR.

The initial two runs were performed using medium resolution spectroscopy with the $K_s$ filter (2.00-2.30 $\mu$m, grism no. 3 with R$\sim$2200, with a dispersion of 4.62~\AA\,pixel$^{-1}$) and a 1$^{\prime\prime}$ slit.
For the remainder of the runs, we switched to the low-resolution spectroscopy with the GRF filter (1.53-2.52 $\mu$m, red grism with R$\sim$980 with a dispersion of 10.22~\AA\,pixel$^{-1}$) and a 0.6$^{\prime\prime}$ slit due to the wider wavelength range it provides, allowing better constraint on the continuum and possible detection of H$\beta$ and H$\gamma$ lines compared with the $K_s$ filter. The $K$-mag values and exposure time of each target are presented in Table \ref{tab:targets}. For the science observations, we used the auto-nod non-destructive readout mode of SOFI provided by the \texttt{SOFI\_spec\_obs\_AutoNodNonDestr} template.
A telluric star was observed after each AGN observation to enable atmospheric correction and flux calibration. 
For the spectral calibration and flat fielding, we took xenon and neon arc lamp observations and dome flat exposures before the start of each night. 
These observations were taken with the \texttt{SOFI\_spec\_cal\_Arcs} and \texttt{SOFI\_spec\_cal\_DomeFlatsNonDestr} observation templates, respectively, using the same slit and filter as our observations for the night.

\section{Data reduction}
\label{sec:data_reduction}

The data were reduced with version 1.5.0 of the  SOFI pipeline.
The flat fields and arc frames were processed using the \texttt{sofi\_spc\_flat} and \texttt{sofi\_spc\_arc} recipes, respectively.
For low-resolution data, we had difficulties obtaining a dispersion solution that matched the $H$- and $K$-bands simultaneously. We, therefore, applied a quadratic correction to each band separately, based on the atmospheric OH lines in the $H$-band and on the arc lines in the $K$-band. 
The science data were processed using \texttt{sofi\_spc\_jitter} recipe to produce a 2D spectrum.
Although this recipe can combine the individual frames and extract a final 1D spectrum, we used our own algorithms for these steps.
Each 2D spectrum was trimmed in the spatial direction, and then a line-by-line residual background was fitted away from the object trace and subtracted.
The frames were aligned to integer pixel precision based on the spectrally summed trace and then combined while rejecting deviant values.
A final iteration of the line-by-line subtraction of the median in each row was then performed on the combined frame.
From this 2D product, we extracted a 1D spectrum on which we performed telluric correction and flux calibration.

The spectral extraction was based on the optimal extraction method described by \cite{Horne1986}, with some adaptations to match it to the pipeline process and the data properties. 
A description of the implementation is given in the Enhanced Resolution Imager and Spectrograph (ERIS)-SPIFFIER Pipeline Manual\footnote{Available from https://www.eso.org/sci/software/pipelines}.
This method is suitable for sources where the spatial distribution changes only gradually with wavelength, including unresolved sources such as stars and the QSOs in our sample.
The routine begins by defining a region around the spectral trace that encompasses all the flux.
This defines the source values $D_{x\lambda}$ as a function of spatial location $x$ and wavelength $\lambda$, and the variance values $V_{x\lambda}$ as the square of the noise.
An initial spectrum is created as $f_\lambda^{initial} = \Sigma_x D_{x\lambda}$ with variance $var[f_\lambda] = \Sigma_x V_{x\lambda}$.
A model of the spatial distribution of the source (or point spread function/PSF) $P_{x\lambda}$ is then constructed by normalising each spectral slice so that $P_{x\lambda} = D_{x\lambda}/f_\lambda$. The resulting model $P_{x\lambda}$ is essentially the probability that a detected photon with wavelength $\lambda$ falls on pixel $x$.
Because $P_{x\lambda}$ is by definition strictly positive, in the first step, any negative values of $P_{x\lambda}$ are set to zero.
The second step is to provide some regularisation along the spectral direction, so at each spatial location $x$, the spectral values of $P_{x\lambda}$ are traced. 
Rather than fit these with low-order polynomials as was done by \cite{Horne1986}, we applied a running median filter. 
In both cases, the same purpose is achieved: to reject outliers, which is the second core part of the algorithm. Any pixel in $P_{x\lambda}$ for which $(D_{x\lambda}-f_\lambda P_{x\lambda})^2 > \sigma_{clip}^2 V$ is set to zero. The threshold $\sigma_{clip}$ is derived using a percentile clipping of the values and is calculated for each spectral row to allow for strong variations in the SNR along the spectrum.
The spectrum estimator is then defined to be a linear combination of unbiased pixel estimates such that 
\[f_\lambda^{unbiased} = \frac{\Sigma_x (W_{x\lambda} D_{x\lambda} / P_{x\lambda})}
                             {\Sigma_x (W_{x\lambda})}
\]
where the variance of the weighted mean is minimised by choosing weights that are inversely proportional to the variance of the variables, so that 
\[
1/W_{x\lambda} = var[D_{x\lambda} / P_{x\lambda}] = V_{x\lambda} / P_{x\lambda}^2
\]
Substituting these weights into the equation above, one can find the optimal extraction of the spectrum $f$ such that when it is multiplied by the source model $P$, the result matches the data $D$, which has variance $V$. The initial optimal spectrum can then be expressed as:
\[
f_\lambda^{optimal} = \frac{\Sigma_x (P_{x\lambda} D_{x\lambda} / V_{x\lambda})}
                         {\Sigma_x (P_{x\lambda}^2 V_{x\lambda})}
\]
with variance
\[
var[f_\lambda^{optimal}] = \frac{1}{\Sigma_x (P_{x\lambda}^2 / V_{x\lambda})}
\]
This process is then iterated a second time replacing the initial estimate of $f_\lambda$ with the first estimate of the optimised spectrum, to yield the final estimate of the optimised spectrum.

The telluric star was used both to correct the atmospheric absorption and for flux calibration.
The former was achieved by modelling the star (spectral type B) as a blackbody with Br$\gamma$ absorption and normalising the resulting telluric spectrum to a maximum value of 1.
The flux calibration was performed by taking the ratio of the counts within 2.0--2.3~$\mu$m and the expected K-band flux calculated from the magnitude. We compare the measured $K_{mag}$ values of our targets based on their average flux densities with the $K_{mag}$ values from their catalogues, and find that our targets are $\sim$0.44 mag fainter than expected. This translates to a $\sim$33\% lower detected flux than expected. We discuss the possible cause of such lower measurement in Sect. \ref{sec:lum_vs_width}.

\begin{figure*}
    \centering
    \includegraphics[width=0.85\textwidth]{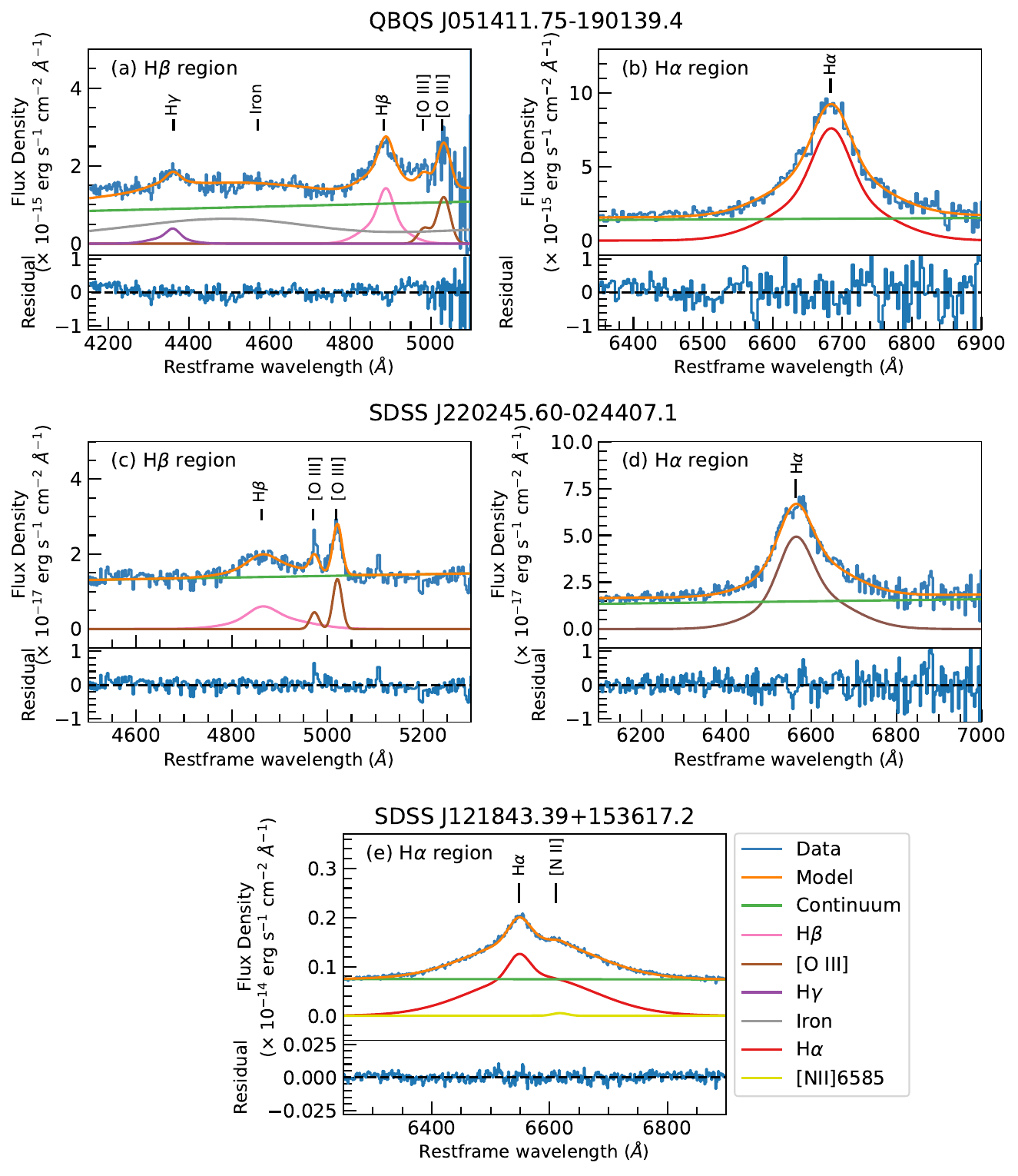}
    \caption{Representative line decomposition results for three of our SOFI $z$ $\sim$ 2 targets. Each row refers to a different target (QBQS J051411.75-190139.4, SDSS J220245.60-024407.1, and SDSS J121843.39+153617.2), while each column refers to a different spectral region (H$\beta$ and H$\alpha$). The observed data are shown in blue, while the cumulative best-fit spectrum is shown as an orange solid line. The different coloured lines pertain to different components in the line decomposition, as shown in the legend. Some of our targets have strong Fe{\small II} and even clear H$\gamma$ emission, as shown in the H$\beta$ region of QBQS J051411.75-190139.4 (panel a). On the other hand, a few targets have prominent (SNR $>$ 3) [O{\small III}] features similar to SDSS J220245.60-024407.1 (panel c). Both of these objects were observed with the GRF filter, hence other Balmer lines aside from H$\alpha$ are detected. On the other hand, SDSS J121843.39+153617.2 was observed with the $K_s$ filter (panel e). Its H$\alpha$ spectrum shows a weak [N{\small II}]$\lambda$6584  feature (see Sect  \ref{sec:line_fitting}).}
    \label{fig:line_decom}
\end{figure*}

\section{Emission line properties}
\label{sec:emission_line_prop}

This section focusses on the observed properties of the emission lines.
We first apply a line decomposition in order to separate the broad component from other features in the spectrum.
We then assess the properties of the broad line emission, in particular, the full-width-half-maximum (FWHM) to $\sigma$ ratios and the H$\alpha$/H$\beta$ flux ratios of our targets.

\subsection{Line decomposition}
\label{sec:line_decom}

In this Section, we describe the fitting of the H$\alpha$, H$\beta$, H$\gamma$, and [O{\small III}] lines as well as the continuum and iron complex (noting that for the medium resolution data taken with the $K_s$-band filter, only the H$\alpha$ line is covered). We used the SAGAN code\footnote{\href {https://github.com/jyshangguan/SAGAN}{https://github.com/jyshangguan/SAGAN}} to decompose the spectra.

The H$\alpha$ lines were fitted with two Gaussian components when there was a clear superposition of a narrower core (that is much broader than the typical width of a narrow-line component expected from the narrow-line region or NLR) and a broader wing component. We impose criteria to determine whether the double Gaussian model is a better fit than the single Gaussian model. Once all criteria are met, the double Gaussian model is preferred. Otherwise, we follow the single Gaussian model fit. Our criteria are similar to that of \citet{Oh2024}:

\begin{enumerate}
    \item The reduced chi-square value of the double Gaussian fitting is less than that of the single Gaussian fitting ($\chi_D^2 < \chi_S^2$), where D and S refer to double and single Gaussian model, respectively;
    \item The SNR of the wing and core components should be greater than 3;
    \item The wing component flux contributes to 10-90 percent of the total flux $f_W/(f_W + f_C)$ where W and C refer to the wing and core components, respectively); and
    \item The velocity dispersion of the broad component should be greater than that of the narrow component by at least its uncertainty ($\sigma_W - \sigma_{W, err} > \sigma_C$).
\end{enumerate}

With these criteria, we were able to confidently choose 21/29 H$\alpha$ lines to be fitted with the double Gaussian model, and the rest were fitted with the single Gaussian model. We also note that most of the emission lines fitted with the single Gaussian model have SNR $\lesssim$ 8. For the H$\beta$ and H$\gamma$ lines, we fit them with the same line profile as the H$\alpha$ line. If the double-Gaussian model is used, the velocity shift difference between the two Gaussian components is also fixed when fitting the other Balmer lines.  We do not assume a theoretical flux ratio of f$_\mathrm{H\alpha}$/f$_\mathrm{H\beta}$ = 3-3.5 as shown by previous works \citep{Dong2008, LaMura2007} as it fails to produce a meaningful fit, which tells us that our targets do not exhibit such a flux ratio. We find three targets that are exempted from our usual line decomposition method: ID\#23, which shows a relatively large deviation between the H$\alpha$ and H$\beta$ velocity shifts ($\sim$1500 km/s) despite being fitted with a similarly shaped Gaussian component; ID\#25, which shows H$\beta$ to be fitted with one Gaussian component while its H$\alpha$ line is fitted with two Gaussian components; and ID\#29, where H$\beta$ is much wider than the H$\alpha$ line. Tables \ref{tab:h-alpha} and \ref{tab:h-beta} show the line-fitting results of H$\alpha$, H$\beta$, [O{\small III}], and H$\gamma$ lines, while Table \ref{tab:h-beta_cont} shows the line-fitting results of the three H$\beta$ lines that did not have the same line profile as their H$\alpha$ lines. We also use the central wavelength of the H$\alpha$ line to verify or update the redshift of each target taken from the source catalogues.

Balmer lines are usually composed of the broad component originating from the BLR, and the narrow component originating from the NLR. Most works use the [{S\small II}] doublet as the narrow line template, but in the case that it is not observable, the [O{\small III}] doublet can be used \citep{Greene2004}. However, the [O{\small III}] doublet suffers from possible contamination of outflowing components, which could indicate a dynamic NLR \citep{Whittle1985, Boroson2005, Marziani2017}, making it a less suitable narrow-line template. To check whether we need to fit the narrow component with a template, we first choose targets with observable [O{\small III}] doublet, i.e. the SNR of [O{\small III}] $>$ 3 and the doublet is not obscured by any atmospheric feature. Seven targets were selected using these criteria. Their Balmer lines were then fitted with the single/double Gaussian model (whichever is suited based on the aforementioned criteria) plus the narrow component with a similar line width as the [O{\small III}] doublet. Afterwards, we measure the narrow components' flux contribution to the fitted Balmer lines. Only three targets (ID \#9, 10, and 17) showed a $>$5\% flux contribution of the narrow components on their Balmer lines. We also do not find any drastic change in our results (i.e. best-fit BLR parameters and virial factor; see Sect. \ref{sec:blr_fitting} and \ref{sec:virial_factor}) and conclusions after removing the fitted narrow components from the Balmer lines of these three targets. Hence, we decided not to include any narrow component fitting in all of our targets.

In most cases, the continuum, which is fit together with the rest of the emission lines, was represented with a power law. However, in a few cases where the power law continuum does not give a good fit (6/29 targets), a 4-degree polynomial was used instead. The polynomial degree was chosen as it is the smallest degree that provides a converging result for these exceptional cases. For these objects fitted with a 4-degree polynomial as their continuum, there is sufficient wavelength range outside the broad line emission.
An iron template based on I Zw 1 \citep{Park2022} was included in the fit when there were clear Fe{\small II} features around the H$\beta$ and [O{\small III}] lines. 
Fig. \ref{fig:line_decom} shows example results of the line decomposition for three targets with different properties: (a) a target observed with the GRF filter which has strong Fe{\small II} features but noisy [O{\small III}] lines due to atmospheric absorption (hence it was not chosen for fitting the narrow component with the [O{\small III}] doublet as a template), (b) a target observed with the GRF filter with slightly asymmetric H$\alpha$, no Fe{\small II} features, very strong [O{\small III}] lines and significant narrow components in their Balmer lines, and (c) a target observed with the $K_s$ filter with a strongly asymmetric H$\alpha$ line that has a bump longwards of the H$\alpha$ central wavelength (and which is discussed in Sect. \ref{sec:line_fitting}).

The uncertainty in the flux density for all spectral channels is calculated as the standard deviation of the fitting residual.
From the decomposition, we measure the line fluxes and luminosities, their FWHM values, and also the dispersions $\sigma$ defined as the square root of the second moment of the line \citep{DallaBonta2020}. Both the FWHM and $\sigma$ are calculated from the best-fit line and are corrected for instrumental broadening.
The 1$\sigma$ uncertainties of these quantities are derived using Monte Carlo techniques, perturbing the spectrum 1000 times with the uncertainty in the flux density. We normalise the Balmer lines by the continuum for BLR fitting (see Sect.~\ref{sec:blr_fitting}).

\subsection{Line properties of the $z\sim2$ targets}
\label{sec:line_fitting}

Among the 29 targets, 24 have data covering both $H$- and $K$- bands.  
Of those, 17 have significant H$\beta$ emission, and two also have observable H$\gamma$ emission. It is important to bear in mind the number of Gaussian components used to fit the Balmer lines. Most of the Balmer emission lines are fitted with two Gaussian components comprising a narrower core component (with $\sigma \lesssim 1200$~km\,s$^{-1}$) and a broad wing component (typically with $1500 \lesssim \sigma \lesssim 3000$~km\,$^{-1}$). 

The corrected redshifts of our targets are almost all consistent within their 1$\sigma$ uncertainties with the redshifts from their respective catalogues, as expected. For 10 of the 17 targets with H$\beta$ emission lines, the [O{\small III}] doublet has been detected and fitted as well, similarly to SDSS J220245.60-024407.1 (ID\#15). Five of these have $\sigma \gtrsim 1000$~km\,s$^{-1}$ for the [O{\small III}] lines.
In addition, six targets have clear Fe{\small II} signatures in their spectra.
We do not investigate these lines or H$\gamma$ further, and instead we focus our analysis on the stronger Balmer lines H$\alpha$ and H$\beta$.

As noted previously, one particular target, ID\#1 (SDSS~J121843.39+153617.2), has a bump longward of the H$\alpha$ peak (see Fig. \ref{fig:line_decom}). 
Its wavelength corresponds to a velocity offset of $\sim 2300$~km\,s$^{-1}$ with respect to H$\alpha$, but only $\sim 1450$~km\,s$^{-1}$ with respect to [N{\small II}]$\lambda$6584 .
Although [N{\small II}] is a doublet, the other line [N{\small II}]$\lambda$6548 is a factor 3 fainter \citep{Acker1989} and so a corresponding feature would not be detectable. In addition, the calculated velocity offsets should not be taken too seriously due to the uncertainty of the redshift taken from the original catalogue which is $\sigma_z$ $\sim$ 0.01 which translates to $\sim$ 3000 km/s \citep{Onken2023}.
We therefore consider the bump to be associated with [N{\small II}]$\lambda$6548 due to the lower velocity offset. We note, however, that the results of our analyses for this target do not change even if the bump is considered as another H$\alpha$ component. We also do not have any other narrow line present in our spectrum of this target to calculate the redshift.

\subsection{FWHM vs. $\sigma$ and the BLR model}
\label{sec:FWHM_vs_sigma}
 
\begin{figure}
    \centering
    \includegraphics[width=\columnwidth]{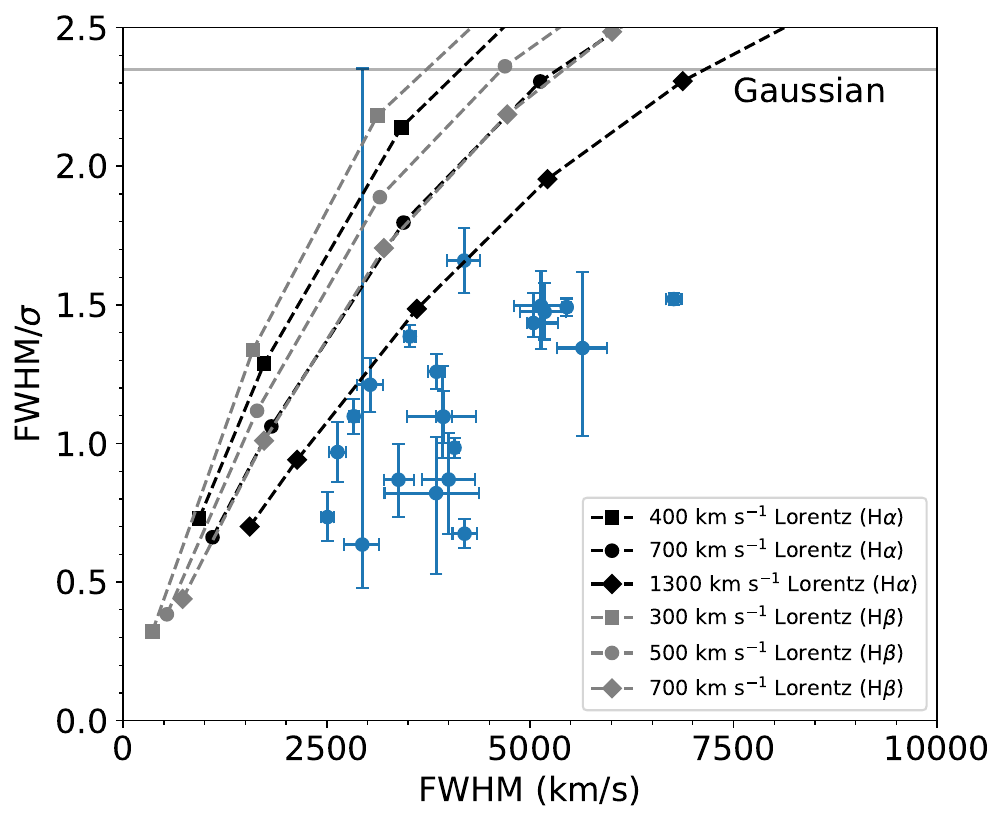}
    \caption{The ratio of FWHM to $\sigma$ (line shape) of H$\alpha$ lines fitted with the double-Gaussian model as a function of FWHM. The Gaussian limit (FWHM $\sim$ 2.35$\sigma$) is shown as a horizontal grey solid line. The error bars denote 1$\sigma$ uncertainties. For comparison, we plot the theoretical line width ratios of rotational line broadened Lorentzian profiles for H$\alpha$ (black dashed lines and markers) and H$\beta$ (grey dashed lines and markers) which were taken from \citet{Kollatschny2011} and \citet{Kollatschny2013}. Different markers pertain to different FWHMs of the Lorentzian profiles.}
    \label{fig:FWHM_ratio}
\end{figure}

One of the most important properties of the line emission in terms of BLR modelling is the shape of the line profile.
A simple way to quantify this is via the ratio of FWHM to $\sigma$, which has been shown to be a good measurement of line shape because FWHM is core sensitive while $\sigma$ is wing sensitive \citep{Wang2019}. We plot the line shape, quantified in this way, as a function of FWHM in Fig.~\ref{fig:FWHM_ratio} for the H$\alpha$ lines. We focus on the H$\alpha$ lines fitted with the double-Gaussian model, as the targets fitted with the single-Gaussian model will lie at the Gaussian limit  (i.e., FWHM/$\sigma = 2.35$) shown as a horizontal line, and we find systematic uncertainty in the H$\beta$ lines (see Sect.~\ref{sec:lum_vs_width}).
Compared with the theoretical line width ratios from \citet{Kollatschny2011} and \citet{Kollatschny2013} (as shown by the black and grey dashed lines), our line width ratios are smaller but show a similar trend as their work and \citet{ Wang2019}: that the FWHM/$\sigma$ increases with FWHM.
Targets with low FWHM and FWHM/$\sigma$ have line profiles that more closely resemble a Lorentzian profile: a superposition of a very broad component and a strong, more prominent core. On the other hand, targets with high FWHM and FWHM/$\sigma$ have line profiles that more closely resemble a Gaussian profile; indeed, at much higher FWHM values, the trend asymptotically reaches the Gaussian limit. 
Comparing our sample with the low redshift work of \citet{Villafana2023}, \citet{Kollatschny2011}, and \citet{Kollatschny2013}, we find no significant difference (Pearson correlation p-value $\gg$ 0.05) in the distribution of line shape versus FWHM between our sample and their AGN samples.

We discuss below two explanations for such profiles. The first scenario, a two-component BLR, would tend to favour fitting the profile with two distinct components. The second scenario, which explains the profile as a combination of turbulence and rotation, would tend to favour fitting the line with a Voigt profile that is a convolution of a Lorentzian profile with a Gaussian.
While we have chosen to fit the profiles with two Gaussians, this is done for the convenience of quantifying $\sigma$, and does not imply a preference for one explanation over the other. 

\subsubsection{Scenario 1: Two-component BLR model}
A two-component BLR has been discussed extensively in the literature \citep[e.g.][]{Brotherton1994, Popovic2004, Zhu2009, Zhang2011, Ludwig2012, Nagoshi2024}. 
In this scenario, the BLR is composed of two components. The first one is an inner disc -- the very broad line region (VBLR) -- that is more closely associated with the accretion disc and is responsible for the broad wings of the observed emission line.
\citet{Zhu2009} suggests that the VBLR represents the ``traditional'' picture of the one-component BLR and is the region responsible for the observed $\sim 0.5$ slope in the size-luminosity relation. 
Indeed, there have been recent claims for detecting Keplerian rotation in this inner disc from the variability of its micro-lensing response \citep{Fian2024}.  
The second component is an outer and more spherical part -- the intermediate line region (ILR) -- which produces the narrow core of the profile. 
The ILR, which is situated at a larger distance from the ionising source, is thought to have higher gas density and be flatter than the VBLR; and is suggested to represent the inner boundary layer between the BLR and the dusty torus of the AGN.
It has been argued that such BLRs with two components occur both in the low-redshift \citep{Zhu2009} and high-redshift \citep{Brotherton1994} Universe. 
\citet{Zhang2011} argued that the ILR of the reverberation-mapped AGN PG~0052+251 was strongly obscured because, in contrast to its H$\alpha$ line, its H$\beta$ line profile does show a clear core component in the line decomposition. While we consider this result uncertain because of the low quality of the spectrum in the H$\beta$ line region of their source, such an explanation could, in principle, apply to the different line widths of H$\alpha$ and H$\beta$ of ID\#15 that was shown in Sect. \ref{sec:line_fitting}. 
\citet{StorchiBergmann2017} also finds that aside from the ubiquity of a disc component in most BLRs, an additional line-emitting component arises at higher Eddington ratios and higher luminosities for Seyfert 1 galaxies. This component is reminiscent of the ILR component, and, as suggested by \citet{StorchiBergmann2017}, may be inflowing \citep{Grier2013}, outflowing \citep{Elitzur2014}, or simply have more elliptical orbits \citep{Pancoast2014}.

The limitation of this explanation is that there is no clear reason why, when considering these two components together, there should be a relation between FWHM and FWHM/$\sigma$ as seen in Fig.~\ref{fig:FWHM_ratio}.
\citet{Collin2006} proposed that this distribution may be associated with the Eddington ratio and, hence, the accretion rate.
They suggested that at large radii, where the self-gravity of the disc overcomes the vertical component of the central gravity due to the SMBH, the resulting cloud collisions due to the gravitational instability would heat the disc and increase its turbulence.
And, based on a correlation between the ratio of the BLR size to this radius and the Eddington ratio, these authors speculated that gravitational instability may be stronger in AGN with higher Eddington ratios, leading to greater turbulence, which perhaps constitutes the start of a disc wind.
This results in a very broad component in the line profile, as exhibited by the low FWHM/$\sigma$ (i.e. low FWHM) sources. On the other hand, weaker accretion produces a more stable BLR, producing Gaussian-like line profiles.

\subsubsection{Scenario 2: Presence of turbulence and rotation}
This relation is specifically addressed in the phenomenological approach put forward by \citet{Kollatschny2011}.
For a disky BLR, the ratio between the turbulent and rotational velocity is proportional to the ratio between the height and radius of the BLR.
Since the rotational velocity is found to increase with increasing FWHM \citep{Kollatschny2013}, objects with high FWHM and, therefore, high FWHM/$\sigma$ have a fast-rotating geometrically thick and flat BLR. In contrast, objects with low FWHM and, therefore, low FWHM/$\sigma$ have slow-rotating spherical BLRs.
These authors noted that while the Balmer lines tend to originate at moderate distances above the disc plane, the highly ionised lines come from smaller radii and at greater scale height and that the resulting geometries resembled disc winds models.
Thus, without explicitly requiring two distinct components, an understanding of the geometry and kinematics of the BLR does lead to insights into the various physical processes occurring in this region.
This perspective matches the approach we adopt when modelling the BLR (see Sect.~\ref{sec:BLR_modelling}). We fit the line profile (and differential phase data from GRAVITY/+ when available) with a single model that encompasses both rotation and dispersion without physically separating them. The dispersion comes directly from the geometry in terms of the thickness, or opening angle, of the BLR and the distribution of clouds within it. It is also affected by whether there is a radial component, whether inflowing or outflow.
Thus, here, too, there is a continuous distribution of potential models from a thin rotating disc through a turbulent, thick rotating disc to a combination of rotation and outflow.

The interpretations above seem likely to be different perspectives on the same underlying processes and geometries that invoke a rotating (thick) disc together with a region or component of that disc where the gas has increased turbulence and scale height, and so may be the origin of the expected disc wind.
However, there is a major difference between them that needs to be resolved.
The wings of the profile trace the rapidly rotating inner disc in the two-component scenario, while they trace the turbulence in the Voigt profile interpretation.
Similarly, the core of the profile traces the outer, more spherical distribution in the two-component model while it traces the rotation in the Voigt profile.
This aspect needs clarification if we are to fully understand the BLR.

\subsection{Luminosity ratio (Balmer decrement)}
\label{sec:lum_vs_width}
As discussed in the previous section, the H$\alpha$ lines are mostly fitted with two Gaussians, while the H$\beta$ and H$\gamma$ lines are fitted with the same line profiles as their respective H$\alpha$ lines with the exception of three targets. To further assess the properties of our targets, we investigate the ratio between H$\alpha$ luminosity and H$\beta$ luminosity (i.e. the Balmer decrement) as a function of H$\alpha$ luminosity.

\begin{figure}
    \centering
    \includegraphics[width=0.9\columnwidth]{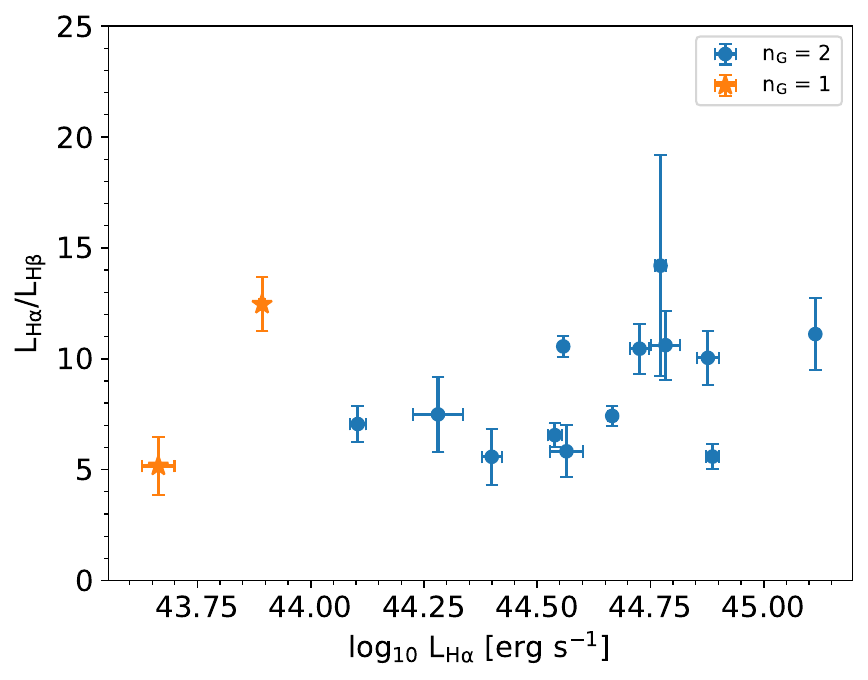}
    \caption{Ratio between H$\alpha$ and H$\beta$ luminosity as a function of H$\alpha$ luminosity. The blue (orange) points show the targets that are fitted with a double (single) Gaussian model. We remove two targets (from the three exceptions in Table \ref{tab:h-beta_cont}) whose H$\beta$ lines cannot be fitted with the same number of Gaussian components and line shape as that of H$\alpha$. The error bars are 1$\sigma$ uncertainties.}
    \label{fig:lum_ratio_vs_Halpha_lum}
\end{figure}

Fig. \ref{fig:lum_ratio_vs_Halpha_lum} shows the Balmer decrement as a function of H$\alpha$ luminosity. The Balmer decrement does not show a significant correlation with L$_\mathrm{H\alpha}$ (the probability of it occurring by chance is $p=0.101$ or 1.3~$\sigma$), although it seems to exhibit a positive correlation (correlation coefficient $\rho$ = 0.41) similar to previous works \citep[e.g.][]{Dominguez2013, Reddy2015}. Furthermore, the Balmer decrements of our targets are $>$ 3.5. The large Balmer decrements suggest that we are missing a significant fraction of the H$\beta$ luminosity and therefore, there is a systematic uncertainty associated with the H$\beta$ emission lines that go beyond the nominal statistical uncertainty derived from the fits. This is due to the limitation of the data rather than having a physical cause, and caution is needed when interpreting values related to H$\beta$, especially the size of the H$\beta$-emitting region of the BLR. 

It is possible that such large Balmer decrements could be due to significant contributions from Wolf-Rayet or late-type OB stars \citep[e.g.][]{Crowther1997}. However, this reason is unlikely to be the cause of the observed Balmer decrements in our sample since these targets are quasar-dominated as per our check of archival spectral data from SDSS DR16Q and UVQS catalogues. The Balmer decrements of our sample are higher than one would expect from typical star-forming galaxies (SFGs) at $z$ $\sim$ 2, which are also found to increase with stellar mass and \citep{Shapley2022, Maheson2024}. Considering also that we find higher H$\alpha$/H$\beta$ ratios than the expected value of $\sim$3.1 \citep{Dong2008, LaMura2007}, these suggest that dust extinction might cause such large Balmer decrements in our sample. To confirm whether dust extinction is the cause of such large H$\alpha$/H$\beta$ ratios in our sample, we estimate the typical extinction coefficient (A$_V$) of our sample based on our median H$\alpha$/H$\beta$ value of $\sim$7.45. Using Eqns. 4 and 7 of \citet{Dominguez2013}, we find A$_V$ $\sim$ 3 which translates to a column density of N$_H$ $\sim$ 6 $\times$ 10$^{21}$ cm$^{-2}$. These values are larger than most SFGs at high-$z$, but still within the acceptable range of A$_V$ values for Type 1-1.5 AGNs \citep{Burtscher2016}. However, such large A$_V$ should also lead to obscuration of the BLR light, which is not the case for our targets. In addition, an extinction coefficient of A$_V$ $\sim$ 3 translates to a detected flux that is $\sim$10\% of the intrinsic flux. However, based on our comparison between our measured and expected $K_{mag}$ values in Sect. \ref{sec:data_reduction}, we find that we are detecting (on average) $\sim$33\% of the intrinsic flux of our sample. Therefore, we cannot conclude with confidence that dust extinction is the root cause of our observed large Balmer decrements.

Nevertheless, other alternative explanations for the large Balmer decrements of AGNs have also been put forward, such as the intrinsic property of the BLR, that is, the BLR consists of clouds with low optical depths and low ionisation parameters \citep{Kwan1981, Canfield1981, Goodrich1990}, or the possible role of accretion rate \citep{Wu2023}. Although we cannot confirm the cause of the large Balmer decrements in our sample, we expect that these should only affect the estimated single-epoch BH masses, BLR radii, and expected differential phase signals of our targets, which are all dependent on the H$\alpha$ luminosity, but not the geometry and virial factors based on our BLR fitting results (see Sect. \ref{sec:BLR_modelling}).

\subsection{BH mass and bolometric luminosity estimation}
\label{sec:Lbol_vs_BHmass}

\begin{figure}
    \centering
    \includegraphics[width=0.9\columnwidth]{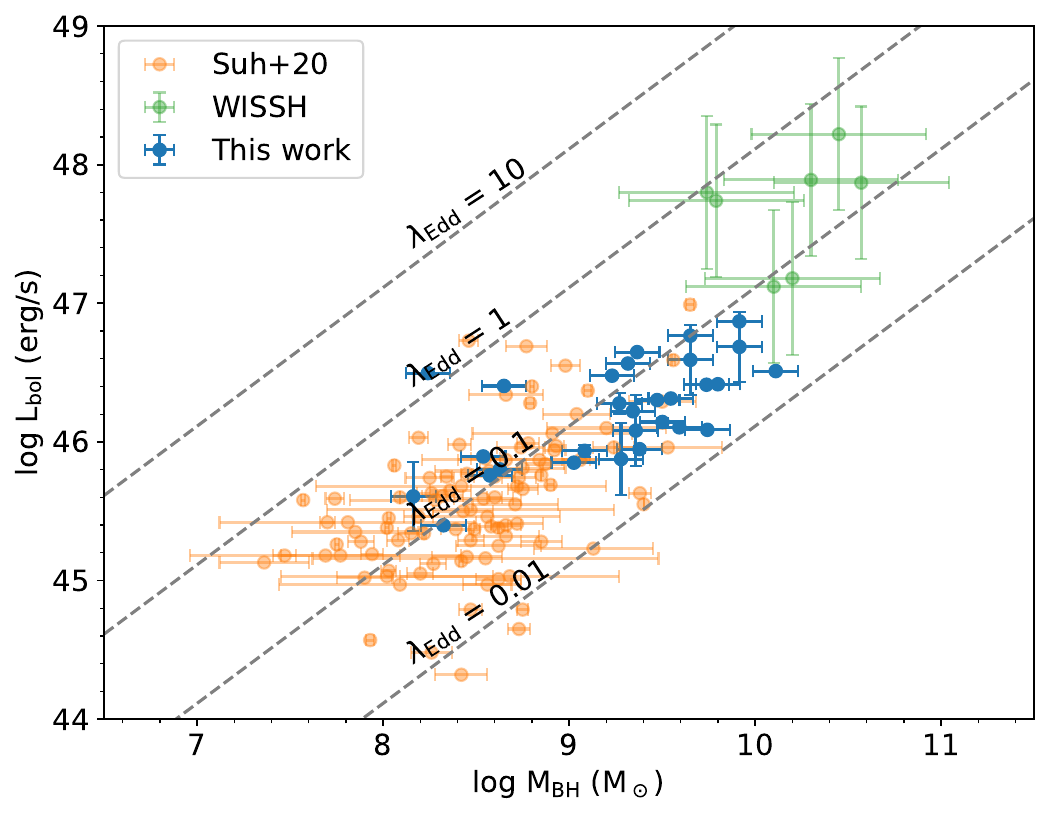}
    \caption{Logarithm of bolometric luminosity as a function of logarithm of black hole mass. For the SOFI $z$ $\sim$ 2 targets (This work, blue data points), the $L_\mathrm{bol}$ are estimated from H$\alpha$, while the $M_\mathrm{BH}$ are estimated using Eqn. 6 of \citet{Woo2015} which uses the dispersion ($\sigma$) and H$\alpha$ luminosity as inputs. The typical error of log $L_\mathrm{bol}$ is shown as the blue vertical error bar on the lower right of the panel. For comparison, we show the sample of $z$ = 1.5 - 2.5 with $L_\mathrm{bol}$ $<$ 47 as orange points \citep{Suh2020} and high-luminosity quasars from the WISSH survey as green points \citep{Bischetti2021}. The grey dashed lines pertain to the loci of the same Eddington ratio: $\lambda_\mathrm{Edd}$ = 0.01, 0.1, 1, and 10. For the WISSH quasars, we assume a BH mass uncertainty of $\sim$ 0.47 dex following the prescription for the 1$\sigma$ relative uncertainty of single-epoch BH mass estimates from \citet{Vestegaard2006}. However, we do not include the systematic uncertainties for the absolute calibration of RM masses.}
    \label{fig:Lbol_vs_Mbh}
\end{figure}

Two of the important parameters we need for comparison with future GRAVITY+ observations of $z$ $\sim$ 2 are SMBH mass ($M_\mathrm{BH}$) and bolometric luminosity ($L_\mathrm{bol}$) estimates. We present the first estimates of $M_\mathrm{BH}$ and $L_\mathrm{bol}$ in columns 7 and 8 of Table \ref{tab:h-alpha}. We present the parameter space that we are probing with our SOFI $z$ $\sim$ 2 targets in Fig. \ref{fig:Lbol_vs_Mbh}. For comparison, we also show the low-luminosity $z$ $\sim$ 2 AGNs from \citet{Suh2020} and the high-luminosity $z$ $\sim$ 2 AGNs from the WISSH survey \citep{Bischetti2021}.

In their Eqns.~5 and~6, \citet{Woo2015} calculated the SMBH mass as a function of H$\alpha$ luminosity and either FWHM or $\sigma$. 
These have different values of the virial factor: for $\sigma$, $f = 4.47$, while for FWHM, $f = 1.12$. 
The resulting SMBH masses of our targets from these equations are consistent with those derived from relations presented elsewhere \citep[e.g.,][]{DallaBonta2020}. 
While there are advantages and disadvantages of different line width measurements for calculating SMBH masses \citep{Peterson2004, Wang2020}, we report the $\sigma$-calculated single-epoch BH mass estimates because it has been argued to have a tighter virial relationship than FWHM \citep{Peterson2004}, and FWHM can lead to overestimation at higher SMBH mass and underestimation at lower mass \citep{DallaBonta2020}. Because not all targets were observed in the necessary band, the $\lambda L_\lambda (5100$~\AA) continuum luminosities are instead calculated using Eqn. 4 of \citet{Woo2015} from the H$\alpha$ luminosity, and the uncertainties are derived by calculating the distribution of SMBH masses via Monte Carlo method, assuming Gaussian distributions of the virial factor $f$, H$\alpha$ luminosities and $\sigma$ values. To convert the $\lambda L_\lambda (5100$~\AA) to $L_\mathrm{bol}$, we used the bolometric correction formula from \citet{Trakhtenbrot2017} which is similar to the bolometric correction of \citet{Suh2020} and \citet{Bischetti2021}. We see that our SOFI $z$ $\sim$ 2 AGNs are located between the two AGN samples, particularly at moderate BH masses (log $M_\mathrm{BH} \sim 8-10.5$) and bolometric luminosities (log $L_\mathrm{bol} \sim 45-47$), which translates to moderate accretion rates (Eddington rates of $\lambda_\mathrm{Edd}$ $\sim$ 0.1).

\section{BLR modelling}
\label{sec:BLR_modelling}

Following the assessment of the line profiles, we now fit them with a BLR model.
In the following subsections, we describe the model used and the simplifications we adopt, the general results, and implications for the virial factor. We finish by looking at two targets for which the asymmetry of the line profiles warrants a more detailed approach than the majority of the sample.

\subsection{Description of the BLR model}
\label{subsec:BLRmodel}

We follow the BLR model fitting methodology introduced by \citet{Kuhn2024}, which was developed based on \citet{Pancoast2014} and \cite{Stock2018}. Rather than model the distribution of BLR clouds and calculate their line emission based on photoionisation physics, we model the distribution of line emission directly. As such, the model focusses on geometry and kinematics without considering the absolute flux scaling.
The model is adapted into a Python package called \texttt{DyBEL} which can be used to fit either a single line (only H$\alpha$) or two or more lines simultaneously (e.g. both H$\alpha$ and H$\beta$). In this section, we briefly describe the most important parameters in the model. We highly recommend the reader to refer to \citet{Pancoast2014} and \citet{GRAVITY2020a} for more quantitative details of the BLR parameters used in our model.

The model assumes a large number of non-interacting clouds (or "line emitting entities", not to be confused with actual physical gas clouds) orbiting a central SMBH with mass $M_\mathrm{BH}$. A shifted Gamma function describes the radial distribution of the clouds. It is controlled by three parameters: the mean radius $\mu$ (in this case, the average emissivity-weighted BLR radius), the fractional inner radius $F = R_\mathrm{min}/\mu$ where $R_\mathrm{min}$ is the minimum BLR radius, and the shape parameter $\beta$ which defines the radial cloud distribution: Gaussian ($0 < \beta < 1$), exponential ($\beta = 1$), or heavy-tailed/steep inner profile ($1 < \beta < 2$). 
The inclination angle, $i$, is the angle of the BLR relative to the plane of the sky and is defined such that a face-on geometry is $i = 0^\circ$, while an edge-on geometry is $i = 90^\circ$. 
The opening angle, $\theta_0$ describes the angular thickness of the BLR and gives the BLR model a ``flared disc'' shape. It is defined such that a thin disc has $\theta_0 = 0^\circ$, while a spherical system has $\theta_0 = 90^\circ$. It is the combination of the flared disc shape and the ellipticity of the orbits (see below) that enables our model to fit targets with a variety of FWHM/$\sigma$.

Asymmetries can also be introduced into the model. For example, the vertical distribution of the clouds is assigned by the parameter $\gamma$, which ranges from 1 to 5. The larger $\gamma$ is, the more concentrated the clouds are towards the disc surface. 
Anisotropic emission from each cloud is parameterised by $\kappa$, which describes the weight of each cloud's emission. The value of $\kappa$ ranges from -0.5 to 0.5. If $\kappa > 0$, clouds closer to the observer (i.e. located on the near side of the BLR) have higher weights, and vice versa. 
Midplane obscuration is controlled by the parameter $\xi$, with $\xi = 1$ pertaining to a transparent midplane (i.e. the numbers of observable clouds on both sides of the midplane are equal), and $\xi = 0$ pertaining to a completely opaque midplane (i.e. the emission of the clouds below the midplane cannot be seen). 

The kinematics of the clouds can include not just circular orbits but also elliptical orbits and, hence, radial motion in a way that is still governed by the gravitational potential of the SMBH.
It is done with the parameter $f_\mathrm{ellip}$, which controls the percentage of clouds in circular or bound orbits and unbound orbits that are dominated by radial motion. Each cloud has a radial ($v_\mathrm{r}$) and tangential ($v_\phi$) velocity component to facilitate inflowing and outflowing clouds.
A binary parameter, $f_\mathrm{flow}$ controls the direction of the radial motion: inwards ($f_\mathrm{flow} < 0.5$) or outwards ($f_\mathrm{flow} > 0.5$).
The components $v_\mathrm{r}$ and $v_\phi$ are randomly distributed around a point on an ellipse in the $v_\mathrm{r}$-$v_\phi$ plane.
Clouds on circular orbits, where $v_\mathrm{circ} = \sqrt{GM_\mathrm{BH}/r}$, are at (0, $\pm v_\mathrm{circ}$).
Clouds dominated by radial motion possess highly elongated orbits with a maximum radial velocity equal to the escape velocity $v_\mathrm{esc} = \sqrt{2}v_\mathrm{circ}$, and are located at the point ($\pm v_\mathrm{esc}$, 0).
For these clouds, there is an extra angular parameter $\theta_e = \arctan(|v_\phi|/v_\mathrm{r})$ which defines the angular position of the clouds around that ellipse, with $\theta_e = 0$ denoting where the orbits are unbound.
Lastly, there are two additional parameters which do not contribute to the BLR geometry (and are thus `nuisance' parameters): the peak flux ($f_\mathrm{peak}$) and central wavelength ($\lambda_c$) of the line. The priors of all parameters are adapted from \citet{GRAVITY2020a}.

The model described above is used to fit the line emission distribution of the BLR, excluding any photoionisation physics. Following previous work \citep{Kuhn2024, GRAVITY2020a, GRAVITY2024, Abuter2024}, we used the Python package \texttt{dynesty} \citep{Speagle2020} together with a nested sampling algorithm \citep{Skilling2004} to fit the data. We used 1200 live points with the dynamic nested sampler (\texttt{DynamicNestedSampler}) and the random walk (\texttt{rwalk}) sampling method. The rest of the options in \texttt{dynesty} were set to their default values. Following \citet{Kuhn2024}, a temperature parameter $T$ was also defined. This parameter was set to 16 in order to provide likelihood functions with fewer peaks and, hence, a better estimation of the posterior distributions. The spectrum was normalised by the continuum so that we effectively fit the line-to-continuum ratio (as a function of observed wavelength). We note that this is also used to estimate the expected differential phase signal of the target (see Sect.~\ref{sec:visphi_estimation}).

Using a spectrum of NGC 3783, \citet{Kuhn2024} demonstrated that fitting H$\alpha$, H$\beta$, H$\gamma$, He{\small I}, and Pa$\beta$ lines simultaneously provide tighter constraints on the BLR parameters of NGC~3783 than fitting them separately and that both methods provide consistent geometry with that derived from the RM and GRAVITY data \citep{GRAVITY2021a, GRAVITY2021b, Bentz2021}.
When an object has H$\beta$ and H$\gamma$ line profiles available, we include them in the fit after tying many of their parameters.
In particular, their central wavelengths are tied so that they all shift by the same small amount 
$\epsilon = \lambda_\mathrm{c}/\lambda_\mathrm{air} -1$, where $\lambda_c$ is the theoretical central wavelength of the line, and $\lambda_\mathrm{air}$ is the wavelength measured in air.
There are two exceptions to this where, because of the wavelength calibration method described in Sect.~\ref{sec:data_reduction}, leaving $\epsilon$ untied yielded better results, and these are indicated in column~3 of Table~\ref{tab:dybel}.
In addition, while allowing the BLR radii derived from each line to be free, we tie the shape of their radial profiles using the $\beta$ and $F$ parameters.
All other parameters are tied except $f_\mathrm{peak}$, which we set to be free for all lines.

It should be noted that since we fit only the spectrum, $R_\mathrm{BLR}$ and $M_\mathrm{BH}$ are fully degenerate because the circular velocities $v_\mathrm{circ}$ of the clouds depend on the ratio of $M_\mathrm{BH}$ and $R_\mathrm{BLR}$. Hence, one of them must be fixed during fitting. We fix $M_\mathrm{BH}$ to the values estimated in Sect. \ref{sec:Lbol_vs_BHmass}. 

Finally, for all of our targets, we try two variations of the BLR model: the full model, which fits all the asymmetry parameters ($\gamma$, $\kappa$, $\xi$, $f_\mathrm{flow}$, $f_\mathrm{ellip}$, and $\theta_e$), and the circular model, for which these are fixed to `neutral' values ($\gamma=1$, $\kappa=0$, $\xi=1$, $f_\mathrm{flow}$, and $f_\mathrm{ellip}=1$ so that $\theta_e$ and $f_\mathrm{flow}$ have no impact).
Our results indicate that in most cases, the resulting BLR geometry and kinematics (in particular, the best-fit values of $i$, $\theta_0$, and $R_\mathrm{BLR}$) are fairly similar for both options.
However, some targets are definitely fitted better with the full model due to their asymmetric profiles (see Sect.~\ref{sec:blr_fitting}). These can be identified in Table~\ref{tab:dybel} by the entries for their asymmetry parameters.

\subsection{BLR fitting results}
\label{sec:blr_fitting}

\begin{figure}
    \centering
    \includegraphics[width=\columnwidth]{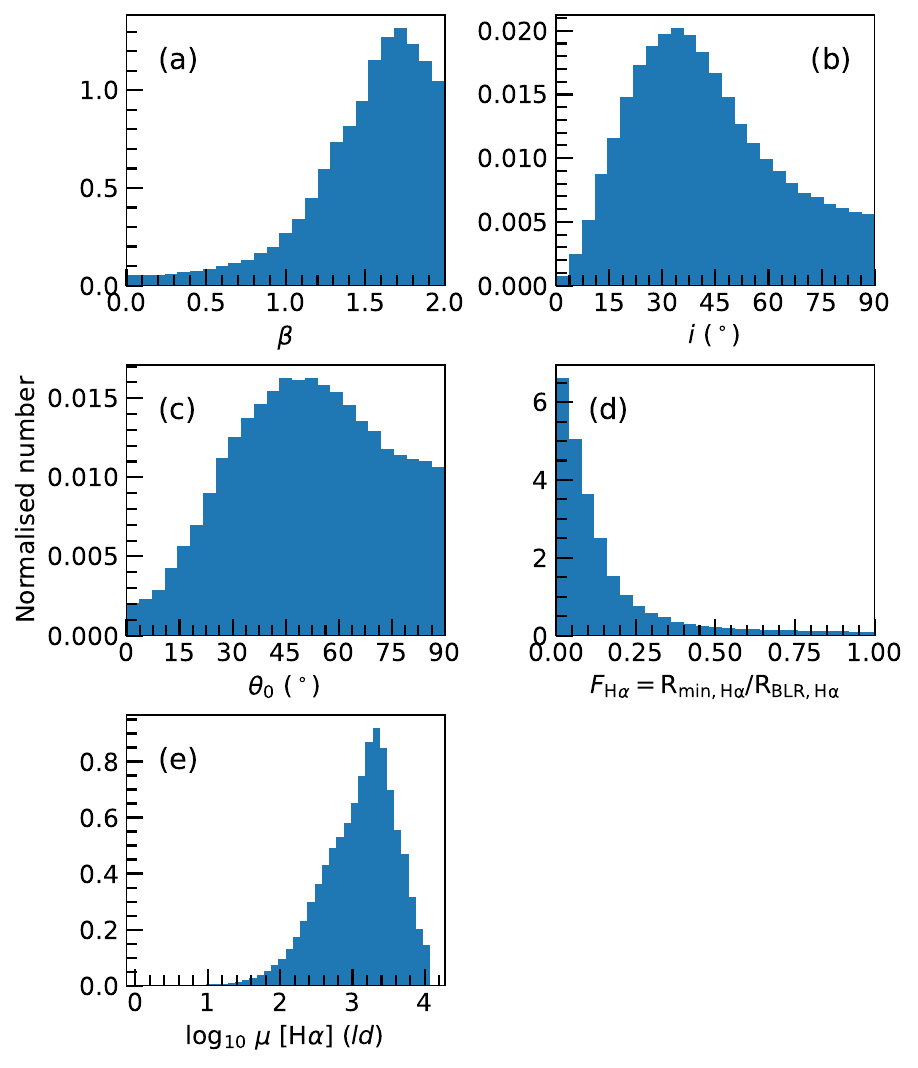}
    \caption{Normalised (i.e. independently for each histogram such that the area under the histogram is 1) histograms showing in blue the summed posterior distributions of the BLR parameters from the best fits to all the $z \sim 2$ targets. The panels correspond to (a) the radial distribution of BLR clouds, (b) the inclination angle, (c) the opening angle, with the minimum and maximum values defining the thin disc and spherical shape, respectively, (d) the ratio between the minimum and mean H$\alpha$ BLR radius, and (e) the mean (emissivity) H$\alpha$ BLR radius.}
    \label{fig:blr_param}
\end{figure}

\begin{figure*}
    \centering
    \includegraphics[width=0.9\textwidth]{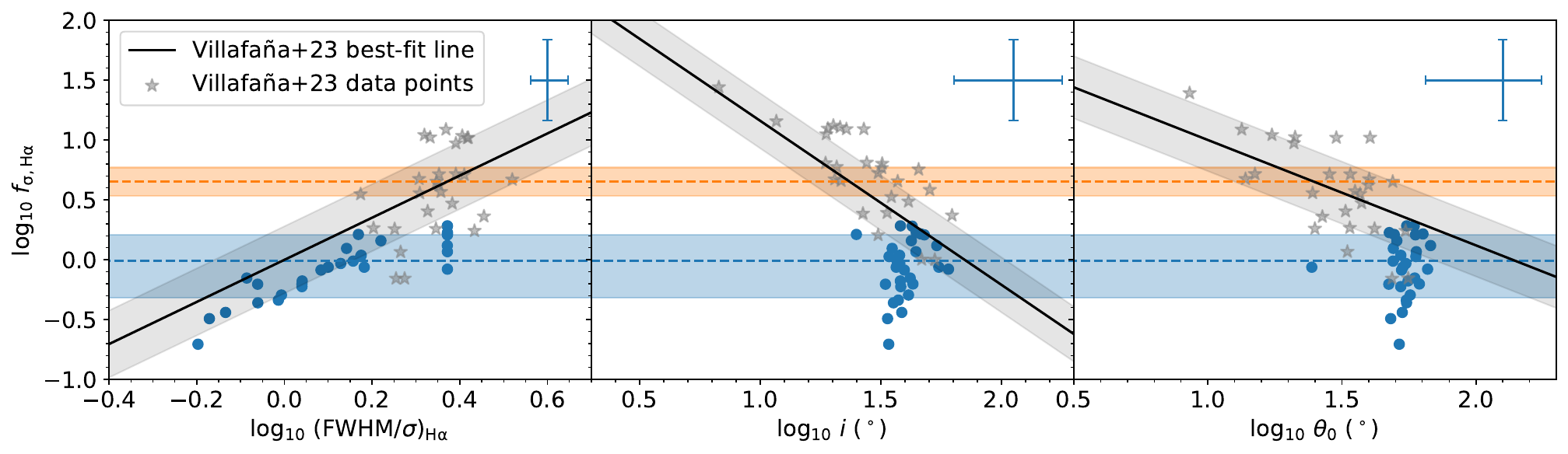}
    \caption{Virial factor $f$$_\sigma$ derived from the \texttt{DyBEL} BLR fitting using $\sigma$-derived SMBH masses, as a function of H$\alpha$ line shape (left panel), inclination angle in degrees (middle panel), and opening angle in degrees (right panel). The average 1$\sigma$ uncertainties are shown on the upper right of each panel.  The orange dashed horizontal line and its 1$\sigma$ range refer to $f = 4.47$, the virial factor in the scaling relations of \citet{Woo2015} from which we derived the SMBH masses to use as input to the fitting procedure. The blue dashed horizontal line and its 1$\sigma$ range refer to the average H$\alpha$ virial factor from our modelling: $\langle f_\mathrm{\sigma, \ H\alpha} \rangle = 1.44$. For comparison, we plot the data points from \citet{Villafana2023} together with a black solid line and a grey region denoting the best-fit relation and its intrinsic scatter.}
    \label{fig:f_vs_shape}
\end{figure*}

In this section, we give an overview of the results from our fits, including the characteristic geometry from the ensemble of best-fit BLR models, and the typical range of values for each fitted parameter. Appendix \ref{tab:dybel} provides the details, listing all the values of the best-fit parameters for each target. 

It is important to note that due to the fact that we are only fitting the spectra of our targets, it is inevitable that our fitting results will yield large uncertainties especially in their best-fit parameters values including those that describe the overall geometry of the BLR such as $\beta$, $i$, and $\theta_0$. Instead of focusing on each individual best-fit parameter values of our targets, we focus on the summed posterior distribution of the best-fit parameters to shed light on the overall behaviour of our fitting results. Therefore, we caution that the individual best-fit values should not be over-interpreted.

Fig.~\ref{fig:blr_param} shows the population distributions for several key BLR parameters ($\beta$, $i$, $\theta_0$, $F_\mathrm{H\alpha}$, and $\mu$). 
Their ranges reflect both the distribution and uncertainty of the individual best-fit values. The distribution for $\beta$ suggests that our targets are typically fitted with a heavy-tailed ($\beta > 1$) radial distribution of BLR clouds with a significant number of line-emitting clouds at larger radii.
The inclination peaking at $i < 45^\circ$ indicates that the BLRs are, as expected, generally viewed closer to face-on than edge-on.
And the opening angle $\theta_0 \sim 50^\circ$ suggests that they tend to have fairly thick discs.
The typical H$\alpha$ BLR radius spans a range from a few hundred to a few thousand light days, which encompasses the radius reported for the $z=2.3$ QSO that was derived from modelling GRAVITY data \citep{Abuter2024} This also reflects the two orders of magnitude range of SMBH mass estimates of our targets as shown in Sect. \ref{sec:Lbol_vs_BHmass}.

\subsection{Virial factor and its dependence on the line shape and BLR parameters}
\label{sec:virial_factor}

The virial factor, which is calculated as $f = GM_\mathrm{BH} / (R_\mathrm{BLR} v^2)$, does not depend on the assumed $M_\mathrm{BH}$ value because the $M_\mathrm{BH}$ the $R_\mathrm{BLR}$ are degenerate (i.e. scaling together without changing the line profile) and $R_\mathrm{BLR}$ is set as a free parameter in our BLR fitting \citep{Kuhn2024}.
Hence, our fits can produce meaningful virial factors for our targets. We note that the choice of using either the circular or full BLR model does not affect the resulting virial factor.
In this Section, we discuss the dependency of the virial factor on various parameters. One focus is on whether one puts $v = \sigma$ or $v = FWHM$.
Another is on $i$ and $\theta_0$, which have been shown to greatly affect the observed line profiles \citep[e.g.,][]{Stock2018, Raimundo2019}
In addition, \citet{Villafana2023} investigated correlations of the virial factor with various parameters -- including those above -- based on 28 low redshift AGNs and dynamical modelling with the same BLR model as \citet{Pancoast2014}. Using the H$\beta$ line, they measured the virial factor for both $\sigma$ and FWHM, as well as for mean and rms spectra. Most of their observed correlations have marginal significance (2-3$\sigma$).
For our analysis, we use the $H\alpha$ line and the BLR size derived from it because of the higher SNR of the $H\alpha$ emission line in our data.

To shed light on this matter, we first calculate $f_\sigma$ using the dispersion of the H$\alpha$ line profile.
Fig. \ref{fig:f_vs_shape} shows the distribution of $f_\sigma$ as a function of H$\alpha$ line shape (FWHM/$\sigma$), inclination angle $i$, and opening angle $\theta_0$. 
We overplot the best-fit lines and data points from \citet{Villafana2023} to compare our results with their work, noting that they used the H$\beta$ line measured in low redshift AGN.
There are several takeaway points we can deduce from Fig.~\ref{fig:f_vs_shape}: 
(1) On the leftmost panel, while our virial factors seem to increase with FWHM/$\sigma$, this is not a significant trend ($p \sim 0.18$). This matches what \citet{Villafana2023} found, and that their steeper trend was not significant when using the mean spectrum, although there was marginal significance for the rms spectrum.
For this comparison, it is important to keep in mind that our sample extends to lower values of line shape to log$_{10} (FWHM/\sigma)_\mathrm{H\alpha} \sim -0.3$.
(2) Our sample also probes larger values of $i$ and $\theta_0$, as seen in the middle and right panels of the figure. While our data do not show any significant correlation with these parameters ($p > 0.40$), they are consistent with the trends of decreasing $f_\sigma$ with increasing $i$ and $\theta_0$ reported by \citet{Villafana2023}.
(3) The quantities in Fig.~\ref{fig:f_vs_shape} have relatively large errors because, in most cases, we fit only a single line profile. As such, it is to be expected that the fitted parameters and their derived quantities will be more uncertain compared to cases where multiple lines are fit \citep{Kuhn2024}.
(4) The average virial factor $f_\sigma=1.44$ we derive (shown as the blue horizontal dashed line in Fig.~\ref{fig:f_vs_shape}) is lower than the virial factor $f_\sigma = 4.47$ from \citet{Woo2015} associated with the calculation of the single-epoch SMBH masses that we use as input to our fits. \citet{Collin2006} pointed out that the virial factor differs for different line shapes. For the mean H$\beta$ spectrum, they find $f = 1.5$ for sources with FWHM/$\sigma \lesssim 1.4$. 
In contrast, \citet{Woo2015} found FWHM/$\sigma \sim 2$ for the $H\alpha$ lines in their sample, close to what is expected for a Gaussian profile, and their resulting $f_\sigma$ is correspondingly higher and very different to $f_{FWHM}$.
We surmise that the low virial factors we find are due to the highly non-Gaussian shape of the H$\alpha$ lines with their prominent extended wings.
We conclude that the line profile shape is an important parameter in this context.
If FWHM/$\sigma$ of a target AGN is very different from that of the objects used to define a scaling relation, the inferred SMBH mass may be biassed, as lower virial factors due to highly non-Gaussian line shapes will give lower single-epoch BH masses.
As such, further investigation of these targets is imperative. Future observations with GRAVITY+ will provide us with an independent and direct measure of the SMBH masses and enable us to assess the error caused by the line shape effects, as well as to create scaling relations specifically for objects where the broad line profile has strong non-Gaussian wings.

\subsection{Targets fitted with the full model}
\label{sec:full_model}

\begin{figure}
    \centering
    \includegraphics[width=0.9\columnwidth]{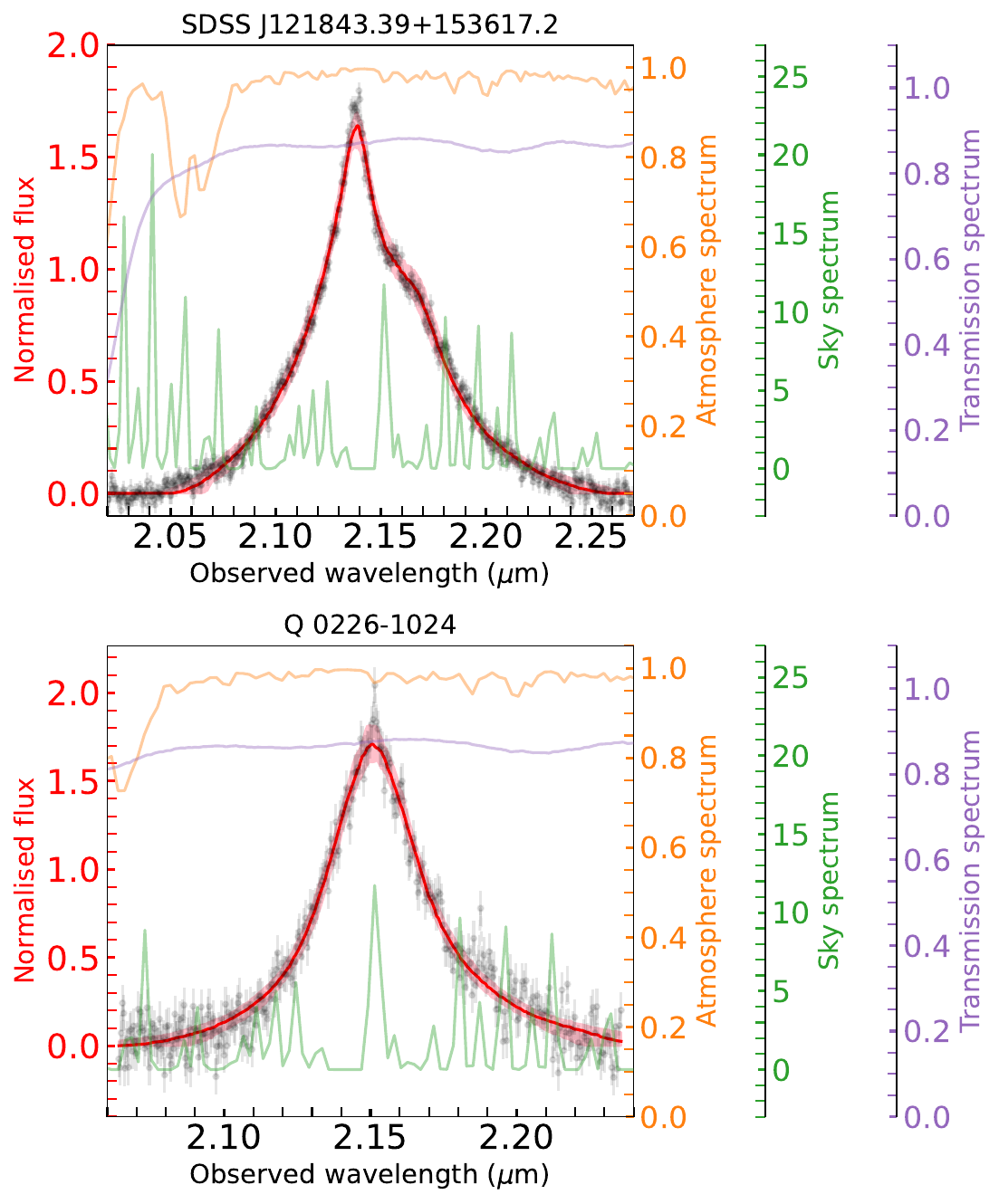}
    \caption{The data (black points) and model (red solid line) H$\alpha$ spectra derived from the best-fit BLR model of two SOFI $z$ $\sim$ 2 targets fitted with the full model. The red-shaded regions show the 1$\sigma$ error of the model spectra. The name of the target is shown on top of each panel. (a) Even after removing the bump on the right side of the H$\alpha$ emission line, which we believe to be [N{\small II}]$\lambda$6584, SDSS J121843.39+153617.2 still shows an asymmetric H$\alpha$ line profile which cannot be fitted with the circular model. (b) The H$\alpha$ emission line of Q 0226-1024 shows an asymmetry in its wings which is better fitted with the full model. We also present the theoretical OH (sky) spectrum, the theoretical atmospheric profile, and the filter transmission profile of SOFI K$_s$ band in orange, green, and purple lines, respectively. The sky and atmospheric profiles are normalised such that the maximum value is 1, and scaled by a factor of 2. We find that the observed asymmetries of both targets are within the high transmission regions of the SOFI K$_s$ filter where no strong sky lines or atmospheric features are present.}
    \label{fig:BLR_full_model}
\end{figure}

Most of our targets (27/29) are well-fitted with the simpler, circular model, indicating that the data currently available -- the line profiles, in particular, their peaks and wings -- are fully consistent with a BLR dominated by Keplerian motion.
This includes some sources with slightly asymmetric profiles, in particular where the wings are offset with respect to the core because the spectra have large enough flux uncertainties that the circular model is still a sufficiently good fit.
However, the asymmetric shape of the line profiles for two of the targets cannot be fitted well with the circular model, partly due to the higher SNR in their spectra.
Instead, for these targets, we use the more complex, full model, which allows anisotropic emission as well as radial motions.
Fig.~\ref{fig:BLR_full_model} shows the spectra and the fitted model profile for these two targets, ID\#1 and ID\#5 (SDSS~J121843.39+153617.2 and Q~0226-1024 respectively).
In the former case, we have removed the small bump that we concluded in Sect.~\ref{sec:line_fitting} was likely due to [N{\small II}]$\lambda$6584.
This still leaves a broad excess on the long wavelength side of the line profile.
If we interpret the small bump as part of the H$\alpha$ line, this only strengthens the results below because it increases the asymmetry.
In ID\#5, although the asymmetry is less obvious, the long wavelength wing is significantly stronger and more extended than the short wavelength wing.

\begin{figure*}
    \centering
    \includegraphics[width=0.9\textwidth]{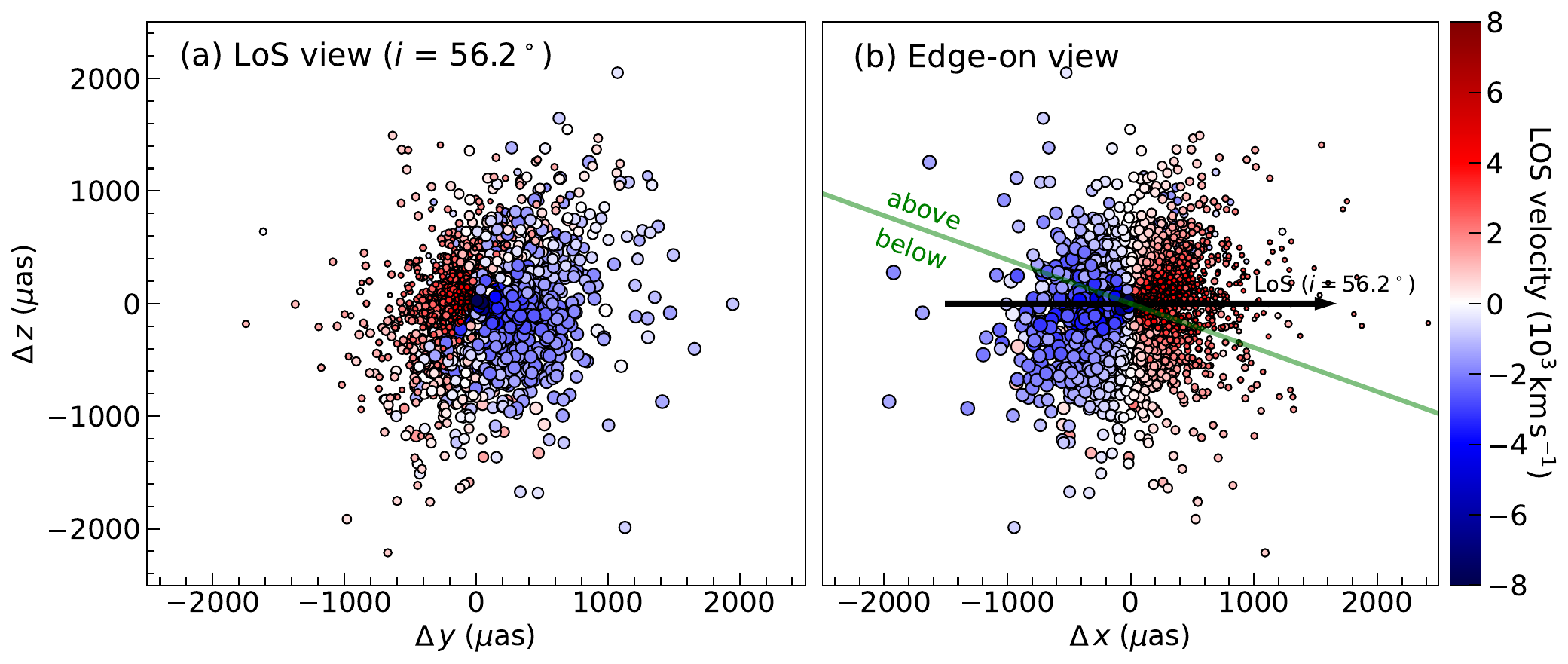}
    \caption{The cloud distribution of the best-fit BLR model of ID\#1 (SDSS J121843.39+153617.2) shown in two different views: (a) line-of-sight (LoS) or face-on view at the best-fit inclination angle $i$ = 54.6$^\circ$ and (b) edge-on view. The PA of the BLR model to generate the cloud distribution is set so that the BLR is perpendicular to the UT4-UT1 baseline to achieve the maximum possible expected differential phase signal on the baseline. The BLR centre is positioned at the origin. The colour of each cloud refers to the LoS velocity, while the size of each cloud refers to the weight of each cloud on the total emission: the larger the size, the greater its contribution to the broad-line emission. The green line on the edge-on view depicts the midplane of the BLR, while the black arrow depicts the LoS of the observer (i.e. the observer is on the +$\Delta$x direction). The LoS is tilted by $i$ which is measured from the line perpendicular to the midplane. Since $i$ $>$ $\theta_0$, the observed flux spectrum is double-peaked. The number of clouds above and below the midplane are the same due to the small midplane opacity. However, the blueshifted clouds have a larger size than the redshifted clouds, indicating the preference of the BLR emission to originate from the far side of the BLR. This explains the relatively strong blueshifted peak of the flux spectrum compared to the redshifted bump.}
    \label{fig:ID1_blr}
\end{figure*}

\begin{figure*}
    \centering
    \includegraphics[width=0.9\textwidth]{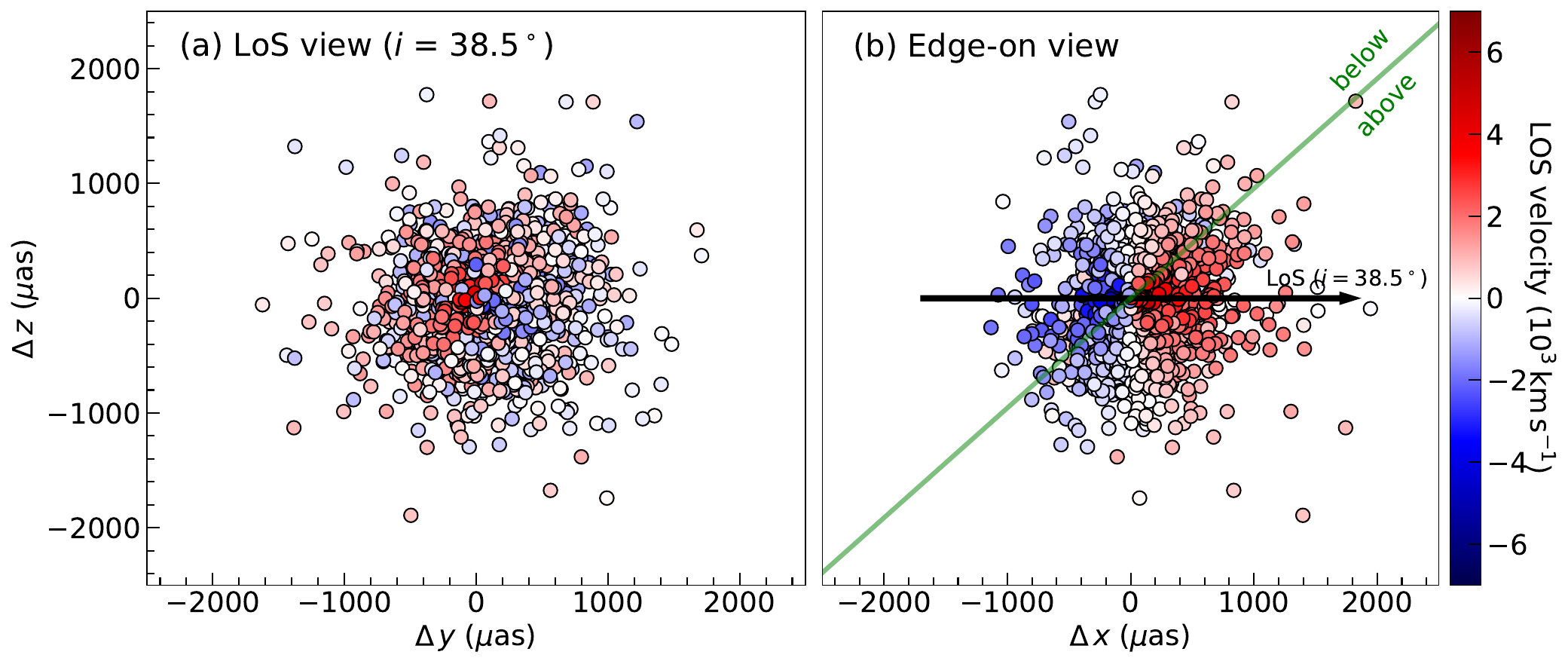}
    \caption{Similar to Fig. \ref{fig:ID1_blr} but for the cloud distribution of the best-fit BLR model of ID\#5 (Q 0226-1024). The BLR clouds have almost equal sizes, pertaining to the lack of preference of the BLR emission to originate from either side. However, the moderate opacity on the midplane affects the blueshifted clouds more than the redshifted clouds, as indicated by the slightly lower number of blueshifted clouds compared to the redshifted clouds. This causes the redshifted wing of the H$\alpha$ emission line to be slightly higher than the blueshifted wing.}
    \label{fig:ID5_blr}
\end{figure*}

In order to assess whether there is a common cause underlying the asymmetric profiles in these objects, we compare their fitted parameters aided by the face-on and edge-on representations of their BLR models in Figs.~\ref{fig:ID1_blr} and~\ref{fig:ID5_blr}. Of the asymmetry parameters, only the midplane transparency is similar for ID\#1 and ID\#5: $\xi \sim 0.5$. This suggests that there is only moderate opacity in the midplane of both their BLRs. While there is little anisotropy in the emission for ID\#5 ($\kappa = -0.01$), for ID\#1 there is a preference for emission from the far side of the BLR ($\kappa = -0.39$). This overcomes the effect of the midplane opacity as can be seen in the edge-on projection of 
the BLR model in Fig.~\ref{fig:ID1_blr}, where the size of the points, which represent clouds, indicates their relative observed flux: although there are slightly fewer points on the far side of the midplane due to its modest opacity, these blue-shifted points are larger than the red-shifted points, hence the former are brighter than the latter.

In terms of the geometry of the BLR, the angular distribution of the clouds for ID\#1 is more concentrated towards the edges ($\gamma = 3.9$) than that of ID\#5 ($\gamma = 2.7$).
In addition, the model for ID\#1 is dominated by radially moving clouds ($f_\mathrm{ellip}$ = 0.17) while that for ID\#5 is more evenly shared between circular and elliptical/radial orbits ($f_\mathrm{ellip}$ = 0.41).
Nevertheless, in both cases the radial motion is inwards ($f_\mathrm{flow} < 0.5$). Lastly, $\theta_e \sim 20$ and $\sim 36$ for ID\#1 and ID\#5 respectively, suggesting that the elliptical orbits of the former are more elongated and so have higher radial velocities, but in neither case do these reach the maximum velocity allowed by the model.

In conclusion, for ID\#1 we purport that the shape of the profile is due to a combination of projection effects resulting from $i$ and $\theta_0$ combined with the effects induced by the asymmetry parameters. First, since $i$ is greater than $\theta_0$ (54.6$^\circ$ versus 25.1$^\circ$), the observed line profile is double-peaked due to the observed biconical structure of the BLR \citep{Stock2018}. The inflowing motion means that clouds on the far side are blueshifted, which clouds on the near side are redshifted.
In addition, since $\kappa$ indicates a preference for the emission to originate from the far side of the BLR, the emission tracing the blueshifted part of the line profile is stronger than the redshifted side, leading to the two peaks having different strengths.

Compared to this, ID\#5 exhibits less asymmetry in its H$\alpha$ profile. 
The edge-on view of its BLR in Fig.~\ref{fig:ID5_blr} clearly shows that the blueshifted clouds are fewer in number than the redshifted clouds due to the moderate asymmetry affecting the former more than the latter. This results in an enhancement in the redshifted wing of the H$\alpha$ emission line of ID\#5.

\section{Differential phase estimation}
\label{sec:visphi_estimation}

One of the main goals of this work is to estimate the strength of the differential phase signals of our $z\sim2$ AGNs to assess their observability with GRAVITY+. The differential phase is one of the most important observables in interferometric observations of AGNs \citep{GRAVITY2018, GRAVITY2020a, GRAVITY2024, GRAVITY2022}. In our context, it is a spatially resolved kinematic signature; specifically, a measure of the astrometric shift of the photocentre of the BLR line emission with respect to that of the continuum as a function of wavelength. More details about the differential phase are presented in \citet{GRAVITY2020a}. A symmetric rotating BLR is expected to show an S-shape differential phase profile, which has been shown to be the case for several GRAVITY-observed AGNs \citep{GRAVITY2018, GRAVITY2020a, GRAVITY2023}. However, some AGNs exhibit asymmetric differential phase profiles, which are explained in terms of asymmetry in the BLR, often combined with outflow motions in the BLR \citep{GRAVITY2024}. 

We derive the expected differential phase of our targets for H$\alpha$ since this is the only line observable in the $K$ band (at $z$ $\sim$ 2) where GRAVITY operates.
We use the normalised line profile from the best-fit BLR model while adopting the same flux uncertainties as the data, and calculate the differential phase as a function of wavelength across the whole $K$ band:
\begin{equation}
    \Delta \phi_\lambda = -2\pi \frac{f_\lambda}{1 + f_\lambda} \vec{u} \cdot \vec{x}_{\mathrm{BLR}, \lambda},
    \label{eqn:visphi_cont}
\end{equation}
where $\Delta \phi_\lambda$ is the differential phase measured in a wavelength channel $\lambda$, $f_\lambda$ is the normalised flux in that channel, \vec{u} is the \textit{uv} coordinate of the baseline, and $\vec{x}_{\mathrm{BLR},\lambda}$ are the BLR photocentres. We calculate the 1$\sigma$ uncertainty of the peak expected differential phase by randomly drawing values of relevant model parameters from the sampled posterior parameter space created during BLR model fitting. This is done 100 times to produce a distribution where the 16th, 50th, and 84th percentile of the peak expected differential phase is calculated.

The position angle PA (measured east of north) rotates the BLR within the sky plane and greatly influences the orientation of the differential phase signal. In order to make a comparative analysis between our targets, we are only interested in the highest possible peak differential phase for each. As such, we focus on the differential phase for the longest baseline, UT4-UT1, for which it is expected to be strongest. We therefore assume that the \textit{uv} direction of the longest baseline of GRAVITY is parallel to the PA of the BLR model.

\begin{figure}
    \centering
    \includegraphics[width=0.9\columnwidth]{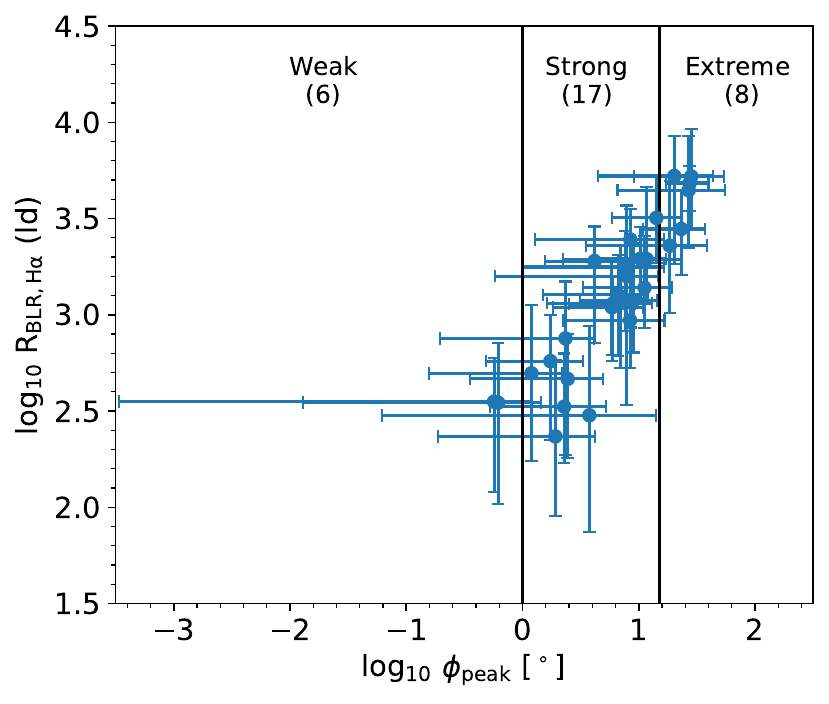}
    \caption{H$\alpha$ BLR size as a function of peak expected differential phase in logarithm. The error bars correspond to their 1$\sigma$ errors. The solid vertical lines pertain to the divisions categorising the targets into weak, strong, and extreme targets (see discussion in Sect. \ref{sec:expected_visphi}). The number of targets for each category is shown in parentheses.}
    \label{fig:BLR_size_vs_visphi}
\end{figure}

\subsection{Expected differential phase signals and effect of fixed BH mass on the \texttt{DyBEL} BLR fitting}
\label{sec:expected_visphi}

Estimating the expected differential phase as described above will provide important guidance in assigning priorities for observations with GRAVITY+, especially for a large sample of AGN. Here, we assess the effectiveness of such a method for our $z\sim2$ targets, together with its caveats from the assumptions made during the BLR fitting.

Fig.~\ref{fig:BLR_size_vs_visphi} shows the H$\alpha$ BLR size of our 29 targets as a function of the peak expected differential phase ($\phi_\mathrm{peak}$) in logarithm. The peak expected differential phase values of all the targets and their 1$\sigma$ errors are listed in Table~\ref{tab:dybel}. The targets are divided into three categories: weak ($\phi_\mathrm{peak} <$ 1$^\circ$), strong (1$^\circ$ $<$ $\phi_\mathrm{peak}$ $<$ 15$^\circ$), and extreme ($\phi_\mathrm{peak}$ $>$ 15$^\circ$. The majority of the targets (23/29) have differential phase signals that go as high as $> 1^\circ$. The strong targets (17/29), have signals with strengths between 1$^\circ-15^\circ$, which is of comparable strength to the $1^\circ$ for the $z=2.3$ QSO observed with GRAVITY \citep{Abuter2024}. On the other hand, the weak targets (6/29) have signals that are comparable to those of the low redshift type~1 AGNs observed with GRAVITY \citep{GRAVITY2018, GRAVITY2020a, GRAVITY2021a}. Most of the targets show a symmetric S-shape signal, as expected for a BLR with ordered rotation. This is a direct result of using the circular model for most sources. In contrast, for the two targets that were fitted with the full model, ID\#1 and ID\#5, we expect an asymmetric differential phase.

The expected differential phase is highly dependent on the assumed (fixed) SMBH mass when fitting the BLR model because the phase signal linearly scales with the BLR size, which is also related to the SMBH mass. However, we expect the SMBH mass not to affect the geometry and the virial factor. Some (6/29) of our targets have very strong differential phase signals ($\ge 15^\circ$). We call these ``extreme'' targets because the SMBH masses derived from their dispersion $\sigma$ are much larger than from their FWHM when using the equations from \citet{Woo2015}. These targets have FWHM/$\sigma < 1.25$, indicating their line wings to be broad and prominent with respect to their narrow cores. Two of these ``extreme'' targets, ID\#9 and ID\#16 (HE~0320-1045 and 2QZ~J031527.8-272645 respectively), show very strong narrow cores, with a narrow core to broad wing amplitude ratio $>10$, while the other ``extreme'' targets have ratios in the range 0.1 to 7.0, similar to the non-extreme targets.

Indeed, \citet{Kollatschny2011} argued that FWHM and $\sigma$ are poor estimators of SMBH mass. They investigated this issue by looking at the turbulent and rotational velocities of several AGNs together with their line shape measurements. Their results suggest that the broadening of the line profiles is due to rotation, and the rotational velocity is a better estimator of SMBH mass than the line dispersion or FWHM.
The reason is that AGNs with similar rotation velocities, for which the SMBH mases are the same, may show different values of FWHM/$\sigma$.
In contrast, the SMBH masses calculated from their FWHM or $\sigma$ may differ drastically. However, inferring the rotational velocity from the line profile in order to estimate the SMBH mass is outside the scope of this work.
An independent and more accurate measurement of the SMBH masses for our targets is crucial to shed light on the size-luminosity relation and the efficacy of the rotation/turbulence interpretation versus the two-component BLR model, underlining the importance of the future GRAVITY+ observations of these targets.

\section{Conclusions and Future Prospects}
\label{sec:conclusion}

To prepare for the advent of GRAVITY+, we performed NTT/SOFI observations of type-1 AGN candidates in order to predict their expected differential phases and assess their priorities for GRAVITY+ observations.
We focus on the 29 $z \sim 2$ targets with prominent H$\alpha$ emission lines.
Among these are 17 for which we have also detected significant H$\beta$ emission, and 2 with H$\gamma$ emission.
We analyse the line profile shape (FWHM/$\sigma$) and fit BLR models using the \texttt{DyBEL} code.
Our results yield the following conclusions:

\begin{enumerate}

\item 
Most of the H$\alpha$ line profiles are highly non-Gaussian and so are fitted with two components: one for the narrow core, and another for the broad wings. This is reminiscent of the two-component BLR model, in which the wings represent an inner fast rotating BLR disc, and the lower velocity core represents an outer thicker part of the BLR. An alternative explanation is that the profile results from a convolution of rotation and turbulence, but in which the rotation is most easily seen via its impact in broadening the core of the profile.

\item 
The average $\sigma$-based H$\alpha$ virial factor of our sample is $f_\sigma \sim 1.44$, which we attribute to the non-Gaussian shape of the emission lines.
In contrast, we would expect to recover $f_\sigma = 4.47$ (as used to derive the single-epoch SMBH masses) if our sample were to exhibit more Gaussian-like line profiles. 

\item
Our sample probes higher inclination $i$ and higher disc thickness $\theta_0$ than those reported by \citet{Villafana2023}, and the values we find are consistent with anti-correlations between these parameters and the virial factor reported by those authors.

\item 
The line profiles of all except two of the targets, are well fitted with a circular simplification of the BLR model. The two targets that require the full model show asymmetry in their H$\alpha$ line profiles, and our results suggest tentative evidence for radially-dominated motions in these targets, with midplane obscuration and anisotropic emission contributing to the asymmetry in the observed line profiles.

\item 
The expected differential phases are an essential tool for assessing the future observing priorities of our targets with GRAVITY+. Among the 29 targets, 23 have strong signals, with 6 possessing expected differential phases $> 15^\circ$. These ``extreme'' targets have very low FWHM/$\sigma$, highlighting concerns about applying scaling relations without accounting for differing line profiles because of the impact this has on the inferred SMBH mass.
GRAVITY+ observations of these targets will provide an independent dynamical measurement of the SMBH in our targets and will not only further our understanding of varying BLR geometries at different epochs and luminosities but will provide a new baseline for future scaling relations.
    
\end{enumerate}

\begin{acknowledgements}
    J.S. acknowledges the National Science Foundation of China (12233001) and the National Key R\&D Program of China (2022YFF0503401). This research has used the NASA/IPAC Extragalactic Database (NED), operated by the California Institute of Technology, under contract with the National Aeronautics and Space Administration. This research used the SIMBAD database operated at CDS, Strasbourg, France.
\end{acknowledgements}

\bibliography{bibliography}{}

\begin{thebibliography}{95}
\expandafter\ifx\csname natexlab\endcsname\relax\def\natexlab#1{#1}\fi

\bibitem[{Abuter {et~al.}(2024)Abuter, Allouche, Amorim, Bailet, Berdeu,
  Berger, Berio, Bigioli, Boebion, Bolzer, \& Bonnet}]{Abuter2024}
Abuter, R., Allouche, F., Amorim, A., {et~al.} 2024, Nature, 627, 281

\bibitem[{Acker {et~al.}(1989)Acker, K\"oppen, Samland, \&
  Stenholm}]{Acker1989}
Acker, A., K\"oppen, J., Samland, M., \& Stenholm, B. 1989, The Messenger, 58,
  44

\bibitem[{Bentz {et~al.}(2013)Bentz, Denney, Grier, Barth, Peterson,
  Vestergaard, Bennert, Canalizo, De~Rosa, Filippenko, Gates, Greene, Li,
  Malkan, Pogge, Stern, Treu, \& Woo}]{Bentz2013}
Bentz, M.~C., Denney, K.~D., Grier, C.~J., {et~al.} 2013, \apj, 767, 149

\bibitem[{Bentz {et~al.}(2021)Bentz, Williams, Street, Onken, Valluri, \&
  Treu}]{Bentz2021}
Bentz, M.~C., Williams, P.~R., Street, R., {et~al.} 2021, \apj, 920, 112

\bibitem[{Bischetti {et~al.}(2021)Bischetti, Feruglio, Piconcelli, Duras, M.,
  Herrero, Venturi, Carniani, Bruni, Gavignaud, \& Testa}]{Bischetti2021}
Bischetti, M., Feruglio, C., Piconcelli, E., {et~al.} 2021, \aap, 645, A33

\bibitem[{Boizelle {et~al.}(2019)Boizelle, Barth, Walsh, Buote, Baker, Darling,
  \& Ho}]{Boizelle2019}
Boizelle, B.~D., Barth, A., Walsh, J.~L., {et~al.} 2019, \apj, 881, 10

\bibitem[{Boroson(2005)}]{Boroson2005}
Boroson, T. 2005, \apj, 130, 381

\bibitem[{Brotherton {et~al.}(1994)Brotherton, Wills, Francis, \&
  Steidel}]{Brotherton1994}
Brotherton, M.~S., Wills, B.~J., Francis, P.~J., \& Steidel, C. 1994, \apj,
  430, 495

\bibitem[{Burtscher {et~al.}(2016)Burtscher, Davies, Graci\'a'-Carpio, Koss,
  Lin, Lutz, Nandra, Netzer, de~Xivry, Ricci, \& Rosario}]{Burtscher2016}
Burtscher, L., Davies, R.~I., Graci\'a'-Carpio, J., {et~al.} 2016, \aap, 586,
  A28

\bibitem[{Canfield \& Puetter(1981)}]{Canfield1981}
Canfield, R.~C. \& Puetter, R. 1981, \apj, 390, 390

\bibitem[{Carraro {et~al.}(2020)Carraro, Rodighiero, Cassata, Brusa, Shankar,
  Baronchelli, Daddi, Delvecchio, Franceschini, Griffiths, \&
  Gruppioni}]{Carraro2020}
Carraro, R., Rodighiero, G., Cassata, P., {et~al.} 2020, \aap, 642, A65

\bibitem[{{Collin} {et~al.}(2006){Collin}, {Kawaguchi}, {Peterson}, \&
  {Vestergaard}}]{Collin2006}
{Collin}, S., {Kawaguchi}, T., {Peterson}, B.~M., \& {Vestergaard}, M. 2006,
  \aap, 456, 75

\bibitem[{Crowther \& Bohannan(1997)}]{Crowther1997}
Crowther, P.~A. \& Bohannan, B. 1997, \aap, 317, 532

\bibitem[{Dalla~Bont\'a {et~al.}(2020)Dalla~Bont\'a, Peterson, Bentz, Brandt,
  Ciroi, De~Rosa, Alvarez, Grier, Hall, Santisteban, \& Ho}]{DallaBonta2020}
Dalla~Bont\'a, E., Peterson, B.~M., Bentz, M.~C., {et~al.} 2020, \apj, 903, 112

\bibitem[{Di~Matteo {et~al.}(2005)Di~Matteo, Springel, \&
  Hernquist}]{DiMatteo2005}
Di~Matteo, T., Springel, V., \& Hernquist, L. 2005, Nature, 433, 604

\bibitem[{Domínguez {et~al.}(2013)Domínguez, Siana, Henry, Scarlata,
  Bedregal, Malkan, Atek, Ross, Colbert, Teplitz, Rafelski, McCarthy, Bunker,
  Hathi, Dressler, Martin, \& D.}]{Dominguez2013}
Domínguez, A., Siana, B., Henry, A.~L., {et~al.} 2013, \apj, 763, 145

\bibitem[{{Dong} {et~al.}(2008){Dong}, {Wang}, {Wang}, {Yuan}, {Zhou}, {Dai},
  \& {Zhang}}]{Dong2008}
{Dong}, X., {Wang}, T., {Wang}, J., {et~al.} 2008, \mnras, 383, 581

\bibitem[{Eisenhauer {et~al.}(2023)Eisenhauer, Monnier, \&
  Pfuhl}]{Eisenhauser2023}
Eisenhauer, F., Monnier, J.~D., \& Pfuhl, O. 2023, \araa, 61, 237

\bibitem[{Elitzur {et~al.}(2014)Elitzur, Ho, \& Trump}]{Elitzur2014}
Elitzur, M., Ho, L.~C., \& Trump, J. 2014, \mnras, 438, 3340

\bibitem[{Ferrarese \& Ford(2005)}]{Ferrarese2005}
Ferrarese, L. \& Ford, H. 2005, Space Sci. Rev., 116, 523

\bibitem[{Ferrarese \& Merritt(2000)}]{Ferrarese2000}
Ferrarese, L. \& Merritt, D. 2000, \apj, 539, L9

\bibitem[{{Fian} {et~al.}(2024){Fian}, {Jim{\'e}nez-Vicente}, {Mediavilla},
  {Mu{\~n}oz}, {Chelouche}, {Kaspi}, \& {For{\'e}s-Toribio}}]{Fian2024}
{Fian}, C., {Jim{\'e}nez-Vicente}, J., {Mediavilla}, E., {et~al.} 2024, \apjl,
  972, L7

\bibitem[{Flesch(2021)}]{Flesch2021}
Flesch, E.~W. 2021, arXiv:2105.12985

\bibitem[{Goodrich(1990)}]{Goodrich1990}
Goodrich, R. 1990, \apj, 355

\bibitem[{{GRAVITY Collaboration et al.}(2017)}]{GRAVITY2017}
{GRAVITY Collaboration et al.} 2017, \aap, 602, A94

\bibitem[{{GRAVITY Collaboration et al.}(2018)}]{GRAVITY2018}
{GRAVITY Collaboration et al.} 2018, \nat, 563, 657

\bibitem[{{GRAVITY Collaboration et al.}(2020a)}]{GRAVITY2020a}
{GRAVITY Collaboration et al.} 2020a, \aap, 643, A154

\bibitem[{{GRAVITY Collaboration et al.}(2021a)}]{GRAVITY2021a}
{GRAVITY Collaboration et al.} 2021a, \aap, 648, A117

\bibitem[{{GRAVITY Collaboration et al.}(2021b)}]{GRAVITY2021b}
{GRAVITY Collaboration et al.} 2021b, \aap, 654, A85

\bibitem[{{GRAVITY+ Collaboration et al.}(2022)}]{GRAVITY2022}
{GRAVITY+ Collaboration et al.} 2022, The Messenger, 189

\bibitem[{{GRAVITY Collaboration et al.}(2023)}]{GRAVITY2023}
{GRAVITY Collaboration et al.} 2023, \aap, 669, A14

\bibitem[{{GRAVITY Collaboration et al.}(2024)}]{GRAVITY2024}
{GRAVITY Collaboration et al.} 2024, A\&A, 684, A167

\bibitem[{Greene \& Ho(2004)}]{Greene2004}
Greene, J.~E. \& Ho, L.~C. 2004, \apj, 610, 722

\bibitem[{Greene {et~al.}(2010)Greene, Peng, Kim, Kuo, Braatz, Impellizzeri,
  Condon, Lo, Henkel, \& Reid}]{Greene2010}
Greene, J.~E., Peng, C.~Y., Kim, M., {et~al.} 2010, \apj, 721, 26

\bibitem[{Grier {et~al.}(2013)Grier, Peterson, Horne, Bentz, Pogge, Denney,
  De~Rosa, Martini, Kochanek, Zu, \& Shappee}]{Grier2013}
Grier, C.~J., Peterson, B.~M., Horne, K., {et~al.} 2013, \apj, 764, 47

\bibitem[{G\"ultekin {et~al.}(2009)G\"ultekin, Richstone, Gebhardt, Lauer,
  Tremaine, Aller, Bender, Dressler, Faber, Filippenko, \&
  Green}]{Gultekin2009}
G\"ultekin, K., Richstone, D.~O., Gebhardt, K., {et~al.} 2009, \apj, 698, 198

\bibitem[{Harrison(2017)}]{Harrison2017}
Harrison, C.~M. 2017, Nat. Astron., 1, 65

\bibitem[{Heckman \& Best(2014)}]{Heckman2014}
Heckman, T. \& Best, P. 2014, \`, 52, 589

\bibitem[{Horne(1986)}]{Horne1986}
Horne, K. 1986, PASP, 98, 609

\bibitem[{{Kaspi} {et~al.}(2021){Kaspi}, {Brandt}, {Maoz}, {Netzer},
  {Schneider}, {Shemmer}, \& {Grier}}]{Kaspi2021}
{Kaspi}, S., {Brandt}, W.~N., {Maoz}, D., {et~al.} 2021, \apj, 915, 129

\bibitem[{Kollatschny \& Zetzl(2011)}]{Kollatschny2011}
Kollatschny, W. \& Zetzl, M. 2011, Nature, 470, 366

\bibitem[{Kollatschny \& Zetzl(2013)}]{Kollatschny2013}
Kollatschny, W. \& Zetzl, M. 2013, \aap, 549, A100

\bibitem[{Kormendy \& Ho(2013)}]{Kormendy2013}
Kormendy, J. \& Ho, L.~C. 2013, \araa, 51, 511

\bibitem[{{Krawczyk} {et~al.}(2013){Krawczyk}, {Richards}, {Mehta}, {Vogeley},
  {Gallagher}, {Leighly}, {Ross}, \& {Schneider}}]{Krawczyk13}
{Krawczyk}, C.~M., {Richards}, G.~T., {Mehta}, S.~S., {et~al.} 2013, \apjs,
  206, 4

\bibitem[{Kuhn {et~al.}(2024)Kuhn, Shangguan, Davies, Man, Cao, Dexter,
  Eisenhauer, F{\"o}rster~Schreiber, Feuchtgruber, Genzel, Gillesen, H{\"o}nig,
  Lutz, Netzer, Ott, Rabien, Santos, Shimizu, Sturm, \& Tacconi}]{Kuhn2024}
Kuhn, L., Shangguan, J., Davies, R., {et~al.} 2024, A\&A, 684, A52

\bibitem[{Kuo {et~al.}(2020)Kuo, Braatz, Impellizzeri, Gao, Pesce, Reid,
  Condon, Kamali, Henkel, \& Greene}]{Kuo2020}
Kuo, C.~Y., Braatz, J.~A., Impellizzeri, C. M.~V., {et~al.} 2020, \mnras, 498,
  1609

\bibitem[{Kwan \& Krolik(1981)}]{Kwan1981}
Kwan, J. \& Krolik, J. 1981, \apj, 250, 478

\bibitem[{{La Mura} {et~al.}(2007){La Mura}, {Popovi{\'c}}, {Ciroi},
  {Rafanelli}, \& {Ili{\'c}}}]{LaMura2007}
{La Mura}, G., {Popovi{\'c}}, L.~{\v{C}}., {Ciroi}, S., {Rafanelli}, P., \&
  {Ili{\'c}}, D. 2007, \apj, 671, 104

\bibitem[{Lapi {et~al.}(2014)Lapi, Raimundo, Aversa, Cai, Negrello, Celotti,
  De~Zotti, \& Danese}]{Lapi2014}
Lapi, A., Raimundo, S., Aversa, R., {et~al.} 2014, \apj, 782, 69

\bibitem[{Li {et~al.}(2023)Li, Li, Shen, Ho, Brandt, Grier, Hall, Homayouni,
  Koekemoer, Schneider, \& Trump}]{Li2023}
Li, J. I.~H., Li, H., Shen, Y., {et~al.} 2023, \apj, 954, 173

\bibitem[{Ludwig {et~al.}(2012)Ludwig, Greene, Barth, \& Ho}]{Ludwig2012}
Ludwig, R.~R., Greene, J.~E., Barth, A.~J., \& Ho, L. 2012, \apj, 756, 51

\bibitem[{Lyke {et~al.}(2020)Lyke, Higley, McLane, Schurhammer, Myers, Ross,
  Dawson, Chabanier, Martini, Des~Bourboux, \& Salvato}]{Lyke2020}
Lyke, B.~W., Higley, A.~N., McLane, J.~N., {et~al.} 2020, \apjs, 250, 8

\bibitem[{Madau \& Dickinson(2014)}]{Madau2014}
Madau, P. \& Dickinson, M. 2014, \araa, 52, 415

\bibitem[{Maheson {et~al.}(2024)Maheson, Maiolino, Curti, Sanders, Tacchella,
  \& Sandles}]{Maheson2024}
Maheson, G., Maiolino, R., Curti, M., {et~al.} 2024, \mnras, 527, 8213

\bibitem[{Marsden {et~al.}(2020)Marsden, Shankar, Ginolfi, \&
  Zubovas}]{Marsden2020}
Marsden, C., Shankar, F., Ginolfi, M., \& Zubovas, K. 2020, Front. Phys., 8, 61

\bibitem[{Mart\'in-Navarro {et~al.}(2020)Mart\'in-Navarro, Burchett, \&
  Mezcua}]{MartinNavarro2020}
Mart\'in-Navarro, I., Burchett, J.~N., \& Mezcua, M. 2020, \mnras, 491, 1311

\bibitem[{Marziani {et~al.}(2017)Marziani, Negrete, Dultzin, Mart\'inez-Aldama,
  Del~Olmo, D'Onofrio, \& Stirpe}]{Marziani2017}
Marziani, P., Negrete, C.~A., Dultzin, D., {et~al.} 2017, Frontiers in Astro.
  and Space Sci., 4, 16

\bibitem[{Monroe {et~al.}(2016)Monroe, Prochaska, Tejos, Worseck, Hennawi,
  Schmidt, Tumlinson, \& Shen}]{Monroe2016}
Monroe, T.~R., Prochaska, J.~X., Tejos, N., {et~al.} 2016, Frontiers in
  Astronomy and Space Sciences, 152, 25

\bibitem[{Nagoshi {et~al.}(2024)Nagoshi, Iwamuro, Yamada, Ueda, Oikawa, Otsuka,
  Isogai, \& Mineshige}]{Nagoshi2024}
Nagoshi, S., Iwamuro, F., Yamada, S., {et~al.} 2024, \mnras, 529, 393

\bibitem[{Oh {et~al.}(2024)Oh, Colless, Barsanti, Zovaro, Croom, Yi, Ristea,
  van~de Sande, D’Eugenio, Bland-Hawthorn, \& Bryant}]{Oh2024}
Oh, S., Colless, M., Barsanti, S., {et~al.} 2024, \mnras, 531, 4017

\bibitem[{Onken {et~al.}(2023)Onken, Wolf, Hon, Lai, Tisserand, \&
  Webster}]{Onken2023}
Onken, C.~A., Wolf, C., Hon, W.~J., {et~al.} 2023, PASA, 40, e010

\bibitem[{Osorno {et~al.}(2023)Osorno, Nagar, Richtler, Humire, Gebhardt, \&
  Gultekin}]{Osorno2023}
Osorno, J., Nagar, N., Richtler, T., {et~al.} 2023, \aap, 679, A37

\bibitem[{Pancoast {et~al.}(2014)Pancoast, Brewer, \& Treu}]{Pancoast2014}
Pancoast, A., Brewer, B.~J., \& Treu, T. 2014, \mnras, 445, 3055

\bibitem[{Park {et~al.}(2022)Park, Barth, Ho, \& Laor}]{Park2022}
Park, D., Barth, A.~J., Ho, L.~C., \& Laor, A. 2022, \apjs, 258, 38

\bibitem[{Peterson(1993)}]{Peterson1993}
Peterson, B.~M. 1993, \pasp, 105, 247

\bibitem[{Peterson(2008)}]{Peterson2008}
Peterson, B.~M. 2008, \apj, 52, 240

\bibitem[{Peterson {et~al.}(2004)Peterson, Ferrarese, Gilbert, Kaspi, Malkan,
  Maoz, Merritt, Netzer, Onken, Pogge, \& Vestergaard}]{Peterson2004}
Peterson, B.~M., Ferrarese, L., Gilbert, K.~M., {et~al.} 2004, \apj, 613, 682

\bibitem[{{Planck Collaboration et al.}(2016)}]{Planck2016}
{Planck Collaboration et al.} 2016, \aap, 594, A13

\bibitem[{Popović {et~al.}(2004)Popović, Mediavilla, Bon, \&
  Ili\'c}]{Popovic2004}
Popović, L.~v., Mediavilla, E., Bon, E., \& Ili\'c, D. 2004, \aap, 423, 909

\bibitem[{Prieto {et~al.}(2022)Prieto, Rodríguez-Ardila, Panda, \&
  Marinello}]{Prieto2022}
Prieto, A., Rodríguez-Ardila, A., Panda, S., \& Marinello, M. 2022, \mnras,
  510, 1010

\bibitem[{Raimundo {et~al.}(2019)Raimundo, Pancoast, Vestergaard, Goad, \&
  Barth}]{Raimundo2019}
Raimundo, S.~I., Pancoast, A., Vestergaard, M., Goad, M.~R., \& Barth, A.~J.
  2019, \mnras, 489, 1899

\bibitem[{Reddy {et~al.}(2015)Reddy, Kriek, Shapley, Freeman, Siana, Coil,
  Mobasher, Price, Sanders, \& Shivaei}]{Reddy2015}
Reddy, N.~A., Kriek, M., Shapley, A.~E., {et~al.} 2015, \apj, 806, 256

\bibitem[{Saglia {et~al.}(2016)Saglia, Opitsch, Erwin, Thomas, Beifiori,
  Fabricius, Mazzalay, Nowak, Rusli, \& Bender}]{Saglia2016}
Saglia, R.~P., Opitsch, M., Erwin, P., {et~al.} 2016, \apj, 818, 47

\bibitem[{Shapley {et~al.}(2022)Shapley, Sanders, Salim, Reddy, Kriek,
  Mobasher, Coil, Siana, Price, Shivaei, \& Dunlop}]{Shapley2022}
Shapley, A.~E., Sanders, R.~L., Salim, S., {et~al.} 2022, \apj, 926, 145

\bibitem[{Shen \& Liu(2012)}]{Shen2012}
Shen, Y. \& Liu, X. 2012, \apj, 753, 125

\bibitem[{Silk(2013)}]{Silk2013}
Silk, J. 2013, \apj, 772, 112

\bibitem[{{Skilling}(2004)}]{Skilling2004}
{Skilling}, J. 2004, in American Institute of Physics Conference Series, Vol.
  735, Bayesian Inference and Maximum Entropy Methods in Science and
  Engineering: 24th International Workshop on Bayesian Inference and Maximum
  Entropy Methods in Science and Engineering, ed. R.~{Fischer}, R.~{Preuss}, \&
  U.~V. {Toussaint}, 395--405

\bibitem[{Speagle(2020)}]{Speagle2020}
Speagle, J.~S. 2020, \mnras, 493, 3132

\bibitem[{Stock(2018)}]{Stock2018}
Stock, M.~R. 2018, Master's thesis, Technische Universität München

\bibitem[{Storchi-Bergmann {et~al.}(2017)Storchi-Bergmann, Schimoia, Peterson,
  Elvis, Denney, Eracleous, \& Nemmen}]{StorchiBergmann2017}
Storchi-Bergmann, T., Schimoia, J. D.~S., Peterson, B.~M., {et~al.} 2017, \apj,
  835, 236

\bibitem[{{Storey-Fisher} {et~al.}(2024){Storey-Fisher}, {Hogg}, {Rix},
  {Eilers}, {Fabbian}, {Blanton}, \& {Alonso}}]{StoreyFisher2024}
{Storey-Fisher}, K., {Hogg}, D.~W., {Rix}, H.-W., {et~al.} 2024, \apj, 964, 69

\bibitem[{Suh {et~al.}(2020)Suh, Civano, Trakhtenbrot, Shankar, Hasinger,
  Sanders, \& Allevato}]{Suh2020}
Suh, H., Civano, F., Trakhtenbrot, B., {et~al.} 2020, \apj, 889, 32

\bibitem[{Tacconi {et~al.}(2020)Tacconi, Genzel, \& Sternberg}]{Tacconi2020}
Tacconi, L.~J., Genzel, R., \& Sternberg, A. 2020, \araa, 58, 157

\bibitem[{Terrazas {et~al.}(2017)Terrazas, Bell, Woo, \&
  Henriques}]{Terrazas2017}
Terrazas, B.~A., Bell, E.~F., Woo, J., \& Henriques, B. 2017, \apj, 844, 170

\bibitem[{Trakhtenbrot {et~al.}(2017)Trakhtenbrot, Ricci, Koss, Schawinski,
  Mushotzky, Ueda, Veilleux, Lamperti, Oh, Treister, \&
  Stern}]{Trakhtenbrot2017}
Trakhtenbrot, B., Ricci, C., Koss, M.~J., {et~al.} 2017, \mnras, 470, 800

\bibitem[{Vestergaard \& Peterson(2006)}]{Vestegaard2006}
Vestergaard, M. \& Peterson, B. 2006, \apj, 641, 689

\bibitem[{Villafa\~{n}a {et~al.}(2023)Villafa\~{n}a, Williams, Treu, Brewer,
  Barth, Bennert, Guo, Bentz, Canalizo, Filippenko, \& Gates}]{Villafana2023}
Villafa\~{n}a, L., Williams, P.~R., Treu, T., {et~al.} 2023, \apj, 948, 95

\bibitem[{Wang {et~al.}(2020)Wang, Shen, Jiang, Grier, Horne, Homayouni,
  Peterson, Trump, Brandt, Hall, \& Ho}]{Wang2020}
Wang, S., Shen, Y., Jiang, L., {et~al.} 2020, \apj, 903, 51

\bibitem[{Wang {et~al.}(2019)Wang, Shen, Jiang, Horne, Brandt, Grier, Ho,
  Homayouni, Jennifer, Li, \& Schneider}]{Wang2019}
Wang, S., Shen, Y., Jiang, L., {et~al.} 2019, \apj, 882, 4

\bibitem[{Whittle(1985)}]{Whittle1985}
Whittle, M. 1985, \apj, 213, 1

\bibitem[{Woo {et~al.}(2015)Woo, Yoon, Park, Park, \& Kim}]{Woo2015}
Woo, J.~H., Yoon, Y., Park, S., Park, D., \& Kim, S.~C. 2015, \apj, 801, 38

\bibitem[{Wu {et~al.}(2023)Wu, Wu, Xue, Lei, \& Lyu}]{Wu2023}
Wu, J., Wu, Q., Xue, H., Lei, W., \& Lyu, B. 2023, \apj, 950, 106

\bibitem[{Yesuf \& Ho(2020)}]{Yesuf2020}
Yesuf, H.~M. \& Ho, L.~C. 2020, \apj, 901, 42

\bibitem[{{Zhang}(2011)}]{Zhang2011}
{Zhang}, X.-G. 2011, \apj, 741, 104

\bibitem[{Zhu {et~al.}(2009)Zhu, Zhang, \& Tang}]{Zhu2009}
Zhu, L., Zhang, S.~N., \& Tang, S. 2009, \apj, 700, 1173

\end{thebibliography}
\bibliographystyle{aa}

\appendix
\onecolumn

\begin{table*}[h!]
\section{Target list and emission line properties}
\label{app:tables}

    \centering
    \caption{All high-redshift targets with clear H$\alpha$ detection from their SOFI spectra.}
    \begin{tabular}{cccccccccc}
    \hline              
    \hline
    ID & Name & $z$ & Date & \makecell{Ave. Seeing \\ ($^{\prime\prime}$)} & Airmass & Filter &\makecell{K-band \\ mag.} & \makecell{Exp. time \\ (s)} & SNR \\
    (1) & (2) & (3) & (4) & (5) & (6) & (7) & (8) & (9) & (10)\\
    \hline
    1 & SDSS J121843.39+153617.2 & 2.266 & \multirow{2}{*}{17 Apr. 2022} &  \multirow{2}{*}{0.71} & 1.42 & Ks & 13.7 & 120 & 45.3 \\
    2 & SDSS J140632.73+091130.4 & 2.165 &  &  & 1.27 &     Ks & 16.0 & 240 & 10.6 \\
    \hline
    3 & SDSS J112521.73+193843.9 & 2.328 & \multirow{2}{*}{20 Apr. 2022} &  \multirow{2}{*}{1.79} & 1.53 & Ks & 16.0 & 240 & 6.1 \\
    4 & SDSS J162449.39+092347.6 & 2.274 & & & 1.41 &  Ks      & 15.2 & 240 & 8.8 \\
    \hline
    5 & Q 0226-1024 & 2.276 & 9 Sept. 2022 & 2.23 & 1.21 &  Ks      & 13.9 & 90 & 18.3 \\
    \hline 
    6 & SDSS J223116.24+224510.8 & 2.367 & 11 Sept. 2022 & 1.43 & 1.69 & GRF & 15.4 & 240 & 39.6 \\
    \hline 
    7 & LAMOSTJ225948.68+052616.1 & 2.178 & 13 Sept. 2022 & 1.98 & 1.84 & GRF & 15.8 & 240 & 8.7 \\
    \hline 
    8 & HE 0037-5155 & 2.127 & 11 Nov. 2022 & 1.96 & 1.11 & GRF & 15.1 & 240 & 21.5 \\
    \hline 
    9 & HE 0320-1045 & 2.282 & \multirow{3}{*}{12 Nov. 2022} & \multirow{3}{*}{1.83} & 1.06 & GRF & 14.5 & 240 & 24.2 \\
    10 & CTS A33.02 & 2.360 & &  & 1.00 & GRF & 12.9 & 90 & 23.2 \\
    11 & SDSS J074556.97+182509.7 & 2.339 & &  & 1.59 & GRF & 15.8 & 240 & 6.4 \\
    \hline 
    12 & SDSS J220607.48+203407.4 & 2.455 & \multirow{2}{*}{13 Nov. 2022} & \multirow{2}{*}{1.91} & 1.59 & GRF & 15.8 & 240 & 9.6 \\
    13 & QBQS J051411.75-190139.4 & 2.609 &  &  & 1.03 & GRF & 15.7 & 240 & 59.7 \\
    \hline 
    14 & SDSS J083713.14+191851.1 & 2.277 & \multirow{4}{*}{14 Nov. 2022} & \multirow{4}{*}{0.93} & 2.01 & GRF & 15.9 & 240 & 4.4 \\
    15 & SDSS J220245.60-024407.1 & 2.403 & & &  1.15 & GRF & 15.6 & 240 & 29.1 \\
    16 & 2QZ J031527.8-272645 & 2.611 & &  & 1.59 & GRF & 15.9 & 240 & 25.6\\
    17 & SDSS J025221.12-085515.6 & 2.296 & &  & 1.07 & GRF & 15.5 & 240 & 11.5 \\
    \hline 
    18 & CT 635 & 2.370 & \multirow{2}{*}{12 Dec. 2022} & \multirow{2}{*}{1.05} & 1.09 & GRF & 15.3 & 240 & 51.2 \\
    19 & SDSS J022819.99-062010.5 & 2.522 & &  & 1.09 & GRF & 15.9 & 240 & 9.6 \\
    \hline 
    20 & SDSS J090938.71+041525.9 & 2.449 & 10 Feb. 2023 & 1.22 & 1.24 & GRF & 15.5 & 240 & 13.2 \\
    \hline 
    21 & LAMOSTJ111209.45+072448.6 & 2.462 & 11 Feb. 2023 & 1.72 & 1.34 & GRF & 15.3 & 240 & 21.1 \\
    \hline 
    22 & 2QZ J113630.3+011949 & 2.138 & \multirow{2}{*}{12 Feb. 2023} & \multirow{2}{*}{0.85} & 1.38 & GRF & 15.7 & 240 & 10.7 \\
    23 & FOCAP QNY4:53 & 2.180 & &  & 1.67 & GRF & 15.8 & 240 & 7.6 \\
    \hline 
    24 & J0504+0055 & 2.340 & \multirow{4}{*}{13 Feb. 2023} & \multirow{4}{*}{1.17} & 1.24 & GRF & 13.0 & 90 & 54.3 \\
    25 & SDSS J093134.31+192622.1 & 2.392 & &  & 1.94 & GRF & 16.0 & 240 & 15.2 \\
    26 & SDSS J122654.39-005430.6 & 2.611 & &  & 1.45 & GRF & 16.1 & 240 & 7.3 \\
    27 & SDSS J153712.90+102557.1 & 2.362 & &  & 2.10 & GRF & 15.9 & 240 & 13.2 \\
    \hline 
    28 & J1006-6246 & 2.320 & \multirow{2}{*}{14 Feb. 2023} & \multirow{2}{*}{1.02} & 1.30 & GRF & 15.5 & 240 & 8.2 \\
    29 & J1315+5206 & 2.145 & &  & 1.40 & GRF & 15.1 & 240 & 12.3 \\
    \hline
    \end{tabular}
    \label{tab:targets}
    {\footnotesize
    \begin{flushleft}
    \textbf{Notes:} 
    The columns show the (1) ID numbers, (2) names, (3) redshifts taken from the catalogues where they are picked from, (4) dates of observation, (5) average seeing, and (6) airmass during observation, (7) used filters for their observations, (8) K-band magnitudes taken from \citet{Flesch2021} and \citet{StoreyFisher2024}, (9) total integration time, and (10) SNR based on the peak line flux of H$\alpha$ line. $K_s$ refers to the $K$ short filter (2.00-2.30 $\mu$m) while GRF refers to the red grism (1.53-2.52 $\mu$m).
    \end{flushleft}}
\end{table*}

\begin{table*}[]
    \caption{Emission line measurements of our high-redshift SOFI targets in the H$\alpha$ spectral region.}
    \label{tab:h-alpha}
    \centering
    \begin{tabular}{cccccccc}
    \hline              
    \hline
    \multicolumn{8}{c}{H$\alpha$} \\
    \hline
    ID & FWHM (km/s) & $\sigma$ (km\,s$^{-1}$) & $z_{\rm corr}$ & \makecell{f$_\mathrm{H\alpha}$ \\ ($\times$ 10$^{-15}$)} & \makecell{No. of \\ Gaussian \\ Comp. (H$\alpha$)} & \makecell{$M_\mathrm{BH}$ \\ $\times$ $10^{9}$ \\ ($M_\odot$)} & \makecell{log$_{10} \ L_\mathrm{bol}$ \\ (erg s$^{-1}$)} \\
    (1) & (2) & (3) & (4) & (5) & (6) & (7) & (8) \\
    \hline
    1 & 6765 & 4448 & 2.269 & 22.90 & 2 & ${8.23}_{-1.99}^{+2.61}$ & 46.69 \\
    2 & 3928 & 3691 & 2.167 & 2.434 & 2 & ${1.90}_{-0.46}^{+0.60}$ & 45.88  \\ 
    3 & 2973 & 1264 & 2.328 & 0.932 & 1 & ${0.15}_{-0.04}^{+0.05}$ & 45.61  \\
    4 & 5179 & 2284 & 2.276 & 3.924 & 2 & ${2.28}_{-0.55}^{+0.72}$ & 46.08 \\
    5 & 5040 & 3500 & 2.276 & 17.27 & 2 & ${4.48}_{-1.08}^{+1.42}$ & 46.59  \\
    6 & 2833 & 2589 & 2.369 & 7.538 & 2 & ${1.70}_{-0.41}^{+0.54}$ & 46.48  \\
    7 & 5135 & 3464 & 2.178 & 5.251 & 2 & ${2.39}_{-0.58}^{+0.76}$ & 45.95  \\
    8 & 3850 & 3067 & 2.132 & 8.015 & 2 & ${2.20}_{-0.53}^{+0.70}$ & 46.22  \\
    9 & 3999 & 4815 & 2.282 & 8.860 & 2 & ${6.30}_{-1.52}^{+2.00}$ & 46.42 \\
    10 & 4195 & 6260 & 2.366 & 12.03 & 2 & ${12.94}_{-3.13}^{+4.11}$ & 46.51 \\
    11 & 4373 & 1858 & 2.341 & 1.682 & 1 & ${0.42}_{-0.10}^{+0.14}$ & 45.80  \\
    12 & 3666 & 1558 & 2.456 & 0.720 & 1 & ${0.21}_{-0.05}^{+0.07}$ & 45.40  \\
    13 & 3521 & 2541 & 2.615 & 7.932 & 2 & ${1.86}_{-0.45}^{+0.59}$ & 46.28  \\
    14 & 4871 & 2069 & 2.276 & 1.235 & 1 & ${0.45}_{-0.45}^{+0.14}$ & 46.40 \\
    15 & 5450 & 3654 & 2.403 & 7.592 & 2 & ${3.52}_{-0.85}^{+1.12}$ & 46.31  \\
    16 & 3937 & 3620 & 2.613 & 4.473 & 2 & ${2.97}_{-0.72}^{+0.94}$ & 46.30  \\
    17 & 5644 & 4436 & 2.299 & 4.462 & 2 & ${3.91}_{-0.94}^{+1.24}$ & 46.11  \\
    18 & 3382 & 4008 & 2.378 & 13.48 & 2 & ${5.47}_{-1.32}^{+1.74}$ & 46.41  \\
    19 & 3845 & 5036 & 2.522 & 4.280 & 2 & ${5.53}_{-1.34}^{+1.76}$ & 46.09  \\
    20 & 4190 & 2529 & 2.455 & 11.93 & 2 & ${2.07}_{-0.50}^{+0.66}$ & 46.57  \\
    21 & 2509 & 3467 & 2.463 & 15.30 & 2 & ${4.48}_{-1.08}^{+1.43}$ & 46.77  \\
    22 & 2633 & 2757 & 2.143 & 3.496 & 2 & ${1.21}_{-0.29}^{+0.39}$ & 45.94  \\
    23 & 4079 & 1733 & 2.182 & 2.082 & 1 & ${0.38}_{-0.09}^{+0.12}$ & 45.76  \\
    24 & 4071 & 4142 & 2.342 & 29.14 & 2 & ${8.22}_{-1.99}^{+2.61}$ & 46.87 \\
    25 & 2938 & 4309 & 2.393 & 2.918 & 2 & ${3.17}_{-0.77}^{+1.01}$ & 46.14  \\
    26 & 5996 & 2547 & 2.612 & 2.310 & 1 & ${1.06}_{-0.26}^{+0.34}$ & 45.85  \\
    27 & 3035 & 2517 & 2.363 & 16.90 & 2 & ${2.34}_{-0.56}^{+0.74}$ & 46.65  \\
    28 & 3160 & 1343 & 2.325 & 1.060 & 1 & ${0.17}_{-0.04}^{+0.06}$ & 46.50  \\
    29 & 4248 & 1805 & 2.422 & 1.504 & 1 & ${0.35}_{-0.08}^{+0.11}$ & 45.90  \\
    \end{tabular}
    {\footnotesize
    \begin{flushleft}
    \textbf{Notes:} 
    The columns show the (1) ID number, (2) FWHM in km/s, (3) line dispersion (square root of the second moment of the line profile or $\sigma$) in km/s, (4) the corrected redshift based on the velocity offset of H$\alpha$ with respect to its theoretical central wavelength, $z_{\rm corr}$, (5) integrated H$\alpha$ flux in ergs cm$^{-2}$ s$^{-1}$, (6) the number of Gaussian components used to fit the H$\alpha$ line profile, (7) single-epoch BH mass estimate using Eqn. 6 of \citet{Woo2015}, and (8) bolometric luminosity using the bolometric correction from \citet{Trakhtenbrot2017} to convert the $\lambda L_\lambda (5100$~\AA) to ${\mathrm L}_\mathrm{bol}$. For ID\#1-5 that were observed with the K$_s$ filter, the $\lambda L_\lambda (5100$~\AA) were estimated from their H$\alpha$ luminosity using  Eqn. 4 of \citet{Woo2015}, and the typical error of their log$_{10}$ $L_\mathrm{bol}$ is $\sim$ 0.25.
    For the rest of the targets that were observed with the GRF filter, the $\lambda L_\lambda (5100$~\AA) were measured via Monte Carlo analysis of best-fit continuum after 1000 instances of fitting, and the typical error of their log$_{10}$ $L_\mathrm{bol}$ is $\sim$ 0.014.
    \end{flushleft}}
\end{table*}

\begin{table*}[]
    \centering
    \caption{Emission line measurements of our high-redshift SOFI targets for H$\beta$, [O{\small III}] doublet, and H$\gamma$ emission lines.}
    \label{tab:h-beta}
    \begin{tabular}{cc|cccc|c}
    \hline              
    \hline
    \multicolumn{2}{c|}{H$\beta$} & \multicolumn{4}{c|}{[O{\small III}]} & \multicolumn{1}{c}{H$\gamma$}\\
    \hline
    ID & \makecell{f$_\mathrm{H\beta}$ \\ ($\times$ 10$^{-15}$)} & FWHM (km\,s$^{-1}$) & $\sigma$ (km\,s$^{-1}$) & \makecell{f$_{5007 \ \AA}$ \\ ($\times$ 10$^{-16}$)} & \makecell{f$_{4959 \ \AA}$\\($\times$ 10$^{-16}$)} & \makecell{f$_\mathrm{H\gamma}$ \\ ($\times$ 10$^{-16}$)} \\
    (1) & (2) & (3) & (4) & (5) & (6) & (7) \\
    \hline
    6 & 1.154 & 5330 & 2269 & 2.996 & 0.9960 & \\
    9 & 1.503 & 1015 & 435 & 2.968 & 0.9866 & \\
    10 & 1.121 & 2998 & 1274 & 5.949 & 1.977 & \\
    13 & 1.068 & 2273 & 967 & 4.987 & 1.657 & 0.1696 \\
    15 & 0.7207 & 1350 & 576 & 3.318 & 1.103 & \\
    16* & 0.7679 & 4880 & 2113 & - & 8.093 & 1.805 \\
    17 & 0.6001 & 1848 & 787 & 1.812 & 0.6024 & \\
    18 & 1.245 & 8679 & 3385 & 3.919 & 1.302 & \\
    20 & 0.8404 & 4190 & 703 & 4.034 & 1.236 & \\
    21 & 3.543 & 1356 & 578 & 3.339 & 1.110 & \\
    22 & 0.503 & & & &  \\
    24 & 2.630 & 6248 & 3254 & 5.963 & 1.978 & \\
    27 & 3.024 & 1183 & 506 & 10.70 & 3.557 & \\
    28 & 0.2048 & & & &  \\
    \end{tabular}
    {\footnotesize
    \begin{flushleft}
    \textbf{Note:} 
    The columns show the (1) ID number (2) integrated H$\beta$ flux, (3) FWHM of [O{\small III}] doublet, (4) line dispersion ($\sigma$) of [O{\small III}] doublet, (5) integrated flux of [O {\small III}]$\lambda$5007 emission line, (6) integrated flux of [O{\small III}]$\lambda$4959 emission line, and (7) integrated H$\gamma$ flux. Each [O{\small III}] line is fitted with one Gaussian component. All fluxes have units of ergs cm$^{-2}$ s$^{-1}$. \\
    *Only [O III] 4959 is fitted.
    \end{flushleft}}
\end{table*}

\begin{table*}[]
    \centering
    \caption{Emission line measurements of three high-redshift SOFI targets in the H$\beta$ region where the H$\alpha$ and H$\beta$ line shapes are not fixed to be similar as discussed in Sec.~\ref{sec:line_decom}.}
    \label{tab:h-beta_cont}
    \begin{tabular}{ccccc}
    \hline              
    \hline
    \multicolumn{5}{c}{H$\beta$ (exceptions)} \\
    \hline
    ID & FWHM (km s$^{-1}$) & $\sigma$ (km s$^{-1}$) & f$_\mathrm{H\beta}$ ($\times$ 10$^{-15}$ ergs cm$^{-2}$ s$^{-1}$) & No. of Gaussian Comp. \\
    (1) & (2) & (3) & (4) & (5) \\
    \hline
    23 & 4079 & 1733 & 0.167 & 1 \\
    25 & 2741 & 1171 & 2.170 & 1 \\
    29 & 5449 & 7987 & 10.15 & 2 \\
    \end{tabular}
\end{table*}

\onecolumn
\begin{landscape}
\section{Summary of \texttt{DyBEL} BLR fitting results}
\label{tab:dybel}
\begin{longtable}{ccccccccccccccc}
\caption{Results of fitting the BLR model to the line profiles.} \\
\hline              
\hline
ID & Model & Tied? & H$\alpha$ $R_\mathrm{BLR}$ [ld] & $\beta$ & \makecell{$\epsilon$ \\ $\times$ 10$^{-3}$} & $i$ $[^\circ$] & $\theta_0$ [$^\circ$] & $\kappa$ & $\gamma$ & $\xi$ & $f_\mathrm{flow}$ & $f_\mathrm{ellip}$ & $\theta_e$ [$^\circ$] & $\Delta\phi_\mathrm{peak}$ \\
(1) & (2) & (3) & (4) & (5) & (6) & (7) & (8) & (9) & (10) & (11) & (12) & (13) & (14) & (15) \\
\hline
\endfirsthead
\caption{continued.}\\
\hline
ID & Model & Tied? & H$\alpha$ $R_\mathrm{BLR}$ [ld] & $\beta$ & \makecell{$\epsilon$ \\ $\times$ 10$^{-3}$} & $i$ $[^\circ$] & $\theta_0$ [$^\circ$] & $\kappa$ & $\gamma$ & $\xi$ & $f_\mathrm{flow}$ & $f_\mathrm{ellip}$ & $\theta_e$ [$^\circ$] & $\Delta\phi_\mathrm{peak}$ \\
(1) & (2) & (3) & (4) & (5) & (6) & (7) & (8) & (9) & (10) & (11) & (12) & (13) & (14) & (15) \\
\hline\hline
\endhead

\hline
\endfoot

\hline
\endlastfoot

1 & F & - & $2442_{-1582}^{+1123}$ & ${1.36}_{-0.21}^{+0.18}$ & ${+4.20}_{-8.27}^{+0.69}$ & ${55}_{-27}^{+23}$ & ${25}_{-6}^{+23}$ & ${-0.39}_{-0.02}^{+0.74}$ & ${3.92}_{-1.81}^{+0.48}$ & ${0.45}_{-0.31}^{+0.33}$ & ${0.43}_{-0.27}^{+0.26}$ & ${0.17}_{-0.08}^{+0.27}$ & ${19}_{-7}^{+35}$ & ${9.3}_{-7.9}^{+7.3}$ \\
2 & C & - & $1128_{-554}^{+457}$ & ${1.86}_{-0.31}^{+0.07}$ & ${-0.34}_{-0.56}^{+0.45}$ & ${43}_{-19}^{+16}$ & ${61}_{-24}^{+17}$ & & & & & & & ${4.6}_{-2.5}^{+4.2}$ \\
3 & C & - & $348_{-229}^{+248}$ & ${1.56}_{-0.92}^{+0.20}$ & ${0.22}_{-1.34}^{+0.74}$ & ${53}_{-27}^{+24}$ & ${67}_{-43}^{+12}$ & & & & & & & ${0.3}_{-0.3}^{+1.2}$ \\
4 & C & - & $597_{-373}^{+400}$ & ${1.83}_{-0.48}^{+0.03}$ & ${-0.12}_{-0.55}^{+0.46}$ & ${25}_{-10}^{+20}$ & ${50}_{-23}^{+18}$ & & & & & & & ${0.9}_{-0.6}^{+1.9}$ \\
5 & F & - & $1899_{-1175}^{+974}$ & ${1.76}_{-0.21}^{+0.13}$ & ${+0.91}_{-1.84}^{+1.09}$ & ${37}_{-20}^{+27}$ & ${49}_{-23}^{+28}$ & ${-0.01}_{-0.34}^{+0.36}$ & ${2.69}_{-1.12}^{+1.48}$ & ${0.48}_{-0.31}^{+0.37}$ & ${0.35}_{-0.24}^{+0.38}$ & ${0.41}_{-0.27}^{+0.34}$ & ${36}_{-25}^{+35}$ & ${3.6}_{-2.7}^{+3.4}$ \\
6 & C & Y & $1552_{-335}^{+1316}$ & ${1.76}_{-0.21}^{+0.13}$ & ${-0.11}_{-0.23}^{+0.21}$ & ${33}_{-9}^{+20}$ & ${50}_{-11}^{+24}$ & & & & & & & ${9.6}_{-4.2}^{+9.4}$ \\
7 & C & - & $905_{-378}^{+1058}$ & ${1.48}_{-0.32}^{+0.35}$ & ${-0.05}_{-0.88}^{+1.07}$ & ${37}_{-14}^{+33}$ & ${53}_{-25}^{+25}$ & & & & & & & ${6.2}_{-4.3}^{+9.5}$ \\
8 & C & - & $1492_{-632}^{+1074}$ & ${1.56}_{-0.28}^{+0.28}$ & ${-0.32}_{-0.55}^{+0.54}$ & ${38}_{-16}^{+23}$ & ${56}_{-21}^{+23}$ & & & & & & & ${7.7}_{-4.9}^{+8.6}$ \\
9 & C & Y & $2931_{-1089}^{+2272}$ & ${1.80}_{-0.28}^{+0.11}$ & ${-0.23}_{-0.36}^{+0.55}$ & ${34}_{-12}^{+22}$ & ${52}_{-17}^{+24}$ & & & & & & & ${25.3}_{-16.4}^{+19.0}$ \\
10 & C & Y & $5199_{-2373}^{+3995}$ & ${1.87}_{-0.29}^{+0.07}$ & ${-0.44}_{-0.68}^{+0.68}$ & ${33}_{-14}^{+18}$ & ${48}_{-17}^{+25}$ & & & & & & & ${21.5}_{-14.0}^{+30.0}$ \\
11 & C & - & $525_{-339}^{+955}$ & ${1.23}_{-0.76}^{+0.50}$ & ${-0.50}_{-2.50}^{+2.93}$ & ${53}_{-6}^{+29}$ & ${44}_{-26}^{+33}$ & & & & & & & ${1.1}_{-0.9}^{+4.2}$ \\
12 & C & - & $252_{-161}^{+338}$ & ${0.94}_{-0.45}^{+0.80}$ & ${-0.06}_{-2.18}^{+1.67}$ & ${37}_{-10}^{+39}$ & ${70}_{-50}^{+7}$ & & & & & & & ${0.9}_{-0.7}^{+2.6}$ \\
13 & C & N & $1354_{-717}^{+1017}$ & ${1.64}_{-0.31}^{+0.22}$ & ${-0.28}_{-0.54}^{+0.37}$ & ${34}_{-16}^{+21}$ & ${53}_{-22}^{+23}$ & & & & & & & ${6.3}_{-3.7}^{+10.2}$ \\
14 & C & - & $330_{-220}^{+423}$ & ${1.39}_{-0.77}^{+0.38}$ & ${-0.14}_{-1.98}^{+2.14}$ & ${48}_{-23}^{+27}$ &${59}_{-40}^{+18}$ & & & & & & & ${1.0}_{-1.0}^{+1.2}$  \\
15 & C & Y & $1273_{-664}^{+776}$ & ${1.68}_{-0.35}^{+0.13}$ & ${0.65}_{-0.80}^{+0.39}$ & ${35}_{-16}^{+28}$ & ${60}_{-29}^{+19}$ & & & & & & & ${6.1}_{-4.1}^{+6.8}$ \\
16 & C & Y & $2339_{-1202}^{+2263}$ & ${1.46}_{-0.27}^{+0.39}$ & ${-0.51}_{-0.80}^{+0.78}$ & ${41}_{-19}^{+27}$ & ${59}_{-29}^{+19}$ & & & & & & & ${8.9}_{-6.1}^{+17.4}$ \\
17 & C & Y & $1151_{-533}^{+718}$ & ${1.44}_{-0.28}^{+0.33}$  & ${-0.53}_{-0.77}^{+0.59}$ & ${39}_{-18}^{+18}$ & ${53}_{-18}^{+25}$ & & & & & & & ${5.8}_{-2.9}^{+5.8}$ \\
18 & C & Y & $2746_{-1140}^{+1645}$ & ${1.79}_{-0.20}^{+0.11}$  & ${-0.07}_{-0.30}^{+0.26}$ & ${32}_{-12}^{+12}$ & ${47}_{-14}^{+18}$ & & & & & & & ${15.6}_{-6.8}^{+16.4}$ \\
19 & C & - & $1072_{-1140}^{+1645}$ & ${1.56}_{-0.98}^{+0.22}$ & ${+0.95}_{-2.93}^{+4.44}$ & ${36}_{-10}^{+40}$ & ${53}_{-32}^{+24}$ & & & & & & & ${3.8}_{-3.1}^{+18.6}$ \\
20 & C & Y & $1150_{-620}^{+1150}$ & ${1.40}_{-0.55}^{+0.40}$ & ${-1.60}_{-1.22}^{+1.20}$ & ${42}_{-20}^{+26}$ & ${51}_{-25}^{+25}$ & & & & & & & ${3.6}_{-2.5}^{+10.6}$ \\
21 & C & Y & $5510_{-3670}^{+2999}$ & ${1.69}_{-0.44}^{+0.18}$ & ${-0.05}_{-0.64}^{+0.51}$ & ${40}_{-23}^{+13}$ & ${55}_{-31}^{+17}$ & & & & & & & ${19.8}_{-12.8}^{+35.1}$ \\
22 & C & Y & $1776_{-946}^{+956}$ & ${1.79}_{-0.19}^{+0.14}$ & ${-0.004}_{-0.292}^{+0.296}$ & ${37}_{-18}^{+22}$ & ${55}_{-23}^{+22}$ & & & & & & & ${12.8}_{-7.8}^{+12.8}$ \\
23 & C & Y & $333_{-162}^{+297}$ & ${1.61}_{-0.45}^{+0.22}$ & ${-0.33}_{-0.55}^{+0.75}$ & ${38}_{-16}^{+31}$ & ${55}_{-25}^{+22}$ & & & & & & & ${1.5}_{-1.1}^{+3.7}$ \\
24 & C & Y &$4651_{-1185}^{+1307}$ & ${1.65}_{-0.11}^{+0.11}$ & ${-0.26}_{-0.19}^{+0.17}$ & ${40}_{-9}^{+9}$ & ${54}_{-10}^{+14}$ & & & & & & & ${25.4}_{-7.8}^{+14.2}$ \\
25 & C & Y & $4252_{-2029}^{+4221}$ & ${1.79}_{-0.49}^{+0.11}$ & ${-0.15}_{-0.66}^{+0.67}$ & ${33}_{-13}^{+26}$ & ${54}_{-26}^{+22}$ & & & & & & & ${21.1}_{-13.3}^{+28.6}$ \\
26 & C & - & $606_{-430}^{+518}$ & ${1.48}_{-0.96}^{+0.26}$ & ${-0.55}_{-2.63}^{+3.29}$  & ${48}_{-22}^{+28}$ & ${61}_{-41}^{+16}$ & & & & & & & ${0.6}_{-0.5}^{+1.7}$ \\
27 & C & Y & $2112_{-1083}^{+2855}$ & ${1.61}_{-0.68}^{+0.21}$ & ${0.08}_{-0.98}^{+0.89}$ & ${38}_{-16}^{+33}$ & ${69}_{-41}^{+9}$ & & & & & & & ${11.1}_{-7.1}^{+35.5}$ \\
28 & C & Y & $351_{-277}^{+515}$ & ${1.56}_{-1.07}^{+0.21}$ & ${-0.02}_{-2.09}^{+1.56}$ & ${35}_{-14}^{+30}$ & ${43}_{-18}^{+33}$ & & & & & & & ${0.3}_{-0.3}^{+1.9}$ \\
29 & C & N & $436_{-256}^{+359}$ & ${1.73}_{-0.52}^{+0.12}$ & ${+23.32}_{-0.73}^{+0.41}$ & ${44}_{-18}^{+33}$ & ${51}_{-42}^{+25}$ & & & & & & & ${1.1}_{-0.8}^{+3.7}$ \\
    \hline

\end{longtable}
{\footnotesize
\begin{flushleft}
    \textbf{Note:} 
    The columns show (1) the ID number, (2) the model used (C for circular, F for full), (3) whether the central wavelength of H$\beta$ and H$\gamma$ was tied to that of H$\alpha$ whenever the other Balmer lines are available, (4) the H$\alpha$ BLR size in ld, (5) the shape parameter of the radial cloud distribution, (6) the amount of shift on the central wavelength of the Balmer line(s), (7) inclination angle, (8) position angle, (9) anisotropy parameter, (10) vertical distribution of clouds, (11) midplane obscuration, (12) binary switch for inflowing ($f_\mathrm{flow}$ $<$ 0.5) and outflowing ($f_\mathrm{flow}$ $>$ 0.5) radial motion, (13) fraction of clouds in circular/bound orbits, (14) angular location for radial orbit distribution, and (15) the peak expected differential phase. All errors are 1$\sigma$ uncertainties. More details about the BLR model parameters are discussed in Sect. \ref{sec:BLR_modelling}, while the calculation of the peak expected differential phase is shown in Sect. \ref{sec:visphi_estimation}.
\end{flushleft}}
\twocolumn
\end{landscape}

\end{document}